%Paper: hep-th/9510241
%From: Shyamoli Chaudhuri <sc@itp.ucsb.edu>
%Date: Tue, 31 Oct 1995 14:58:29 -0800
%Date (revised): Tue, 31 Oct 95 21:28:36 PST

\input harvmac
\global\newcount\fno \global\fno=0
\def\myfoot#1{\global\advance\fno by1
\footnote{$^{\the\fno}$}{#1}}

\def\cmp#1{{\it Comm. Math. Phys.} {\bf #1}}
\def\pl#1{{\it Phys. Lett.} {\bf #1B}}

\def\prd#1{{\it Phys. Rev.} {\bf D#1}}

\def\np#1{{\it Nucl. Phys.} {\bf B#1}}

\def\p{{\scriptscriptstyle +}}
\def\m{{\scriptscriptstyle -}}

\def\darr#1{\raise1.5ex\hbox{$\leftrightarrow$}\mkern-16.5mu #1}
 %pound sterling
\def\half{{1\over2}}

\def\msize{12pt}
\def\b({(\kern -5pt}
\def\e){\kern 2pt)}
\def\ha{
\hbox to\msize{\hfil${{\scriptstyle 1}\over{\scriptstyle 2}}$}}
\def\mha{
\hbox to\msize{\hfil--${{\scriptstyle 1}\over{\scriptstyle 2}}$}}
\def\qt{
\hbox to\msize{\hfil${{\scriptstyle 1}\over{\scriptstyle 4}}$}}
\def\mqt{
\hbox to\msize{\hfil--${{\scriptstyle 1}\over{\scriptstyle 4}}$}}
\def\mtq{
\hbox to\msize{\hfil--${{\scriptstyle 3}\over{\scriptstyle 4}}$}}
\def\ze{\hbox to\msize{\hfil$0$}}
\def\one{\hbox to\msize{\hfil$1$}}
\def\mone{\hbox to\msize{\hfil--$1$}}

\def\roughly#1{\raise.3ex\hbox{$#1$\kern-.75em\lower1ex\hbox{$\sim$}}}

\def\p{\raise2pt\hbox to5pt{\fiverm +}}
\def\m{\raise2pt\hbox to5pt{\fiverm --}}

\def\cmp#1{{\it Comm. Math. Phys.} {\bf #1}}
\def\pl#1{{\it Phys. Lett.} {\bf #1B}}

\def\prd#1{{\it Phys. Rev.} {\bf D#1}}

\def\np#1{{\it Nucl. Phys.} {\bf B#1}}

\def\darr#1{\raise1.5ex\hbox{$\leftrightarrow$}\mkern-16.5mu #1}
 %pound sterling
\def\half{{1\over2}}

 %puts a small sixteenth
\catcode`@=11
\def\myeqalign#1{\null\,\vcenter{\openup-1\jot \m@th
 \ialign{\strut$\displaystyle{##}$&$\displaystyle{{}##}$\hfil
	\crcr#1\crcr}}\,}
%%%%%%%%%%%%%%%%%%%%%%%%%%%%%%
%%%%%%%%%%%%%%%%%%%%%%%%%%%%%%%%%%%%%%%%%%%%%%%%%%%%%%%%%%%%%%%%%%%%
 %puts a small three-half
 %puts a small 1/48 in a displayed eqn
\def\roughly#1{\raise.3ex\hbox{$#1$\kern-.75em\lower1ex\hbox{$\sim$}}}
%%%%%%%%%%%%%%%%%%%%%%%%%%%%%%%%%%%%%%%%%%%%%%%%%%%%%%%%%%%%%%%%%%%%
\lref\cy{P. Candelas, G. Horowitz, A. Strominger and E. Witten,
{\it Nucl. Phys.} {\bf B258} (1985) 46.}

\lref\cyphen{B. Greene, K. Kirklin, P. Miron, and G. Ross,
{\it Nucl. Phys.} {\bf B292} (1987) 606; {\it Nucl. Phys.} {\bf B278}
(1986) 667; B. Greene, {\it Phys. Rev.} {\bf D40} (1989) 1645;
R. Arnowitt and P. Nath, {\it Phys. Rev} {\bf D42} (1990) 2948.}

\lref\gepner{D. Gepner, {\it Nucl. Phys.} {\bf B296 } (1988) 757;
C. Scheich and M. Schmidt, {\it Int. Jour. Mod. Phys.} {\bf A7}
(1992) 8021.}

\lref\rolf{R. Schimmrigk, {\it Phys. Lett.} {\bf 193B}
(1987) 175; D. Gepner, Princeton preprint PUPT-88-0085.}

\lref\wit{E. Witten, {\it Nucl. Phys.}\/ {\bf B268} (1986) 1.}

\lref\bdfm{T. Banks, L. Dixon, D. Friedan, and E. Martinec,
{\it Nucl. Phys.} {\bf B299} (1988) 613.}

\lref\kachru{S. Kachru, {\it Phys. Lett.} {\bf B349} (1995) 76;
E. Silverstein and E. Witten, \np{444} (1995) 161;
J. Distler and S. Kachru,
{\it Nucl. Phys.} {\bf B413} (1994) 213.}

\lref\fernando{L. Ibanez, H. Nilles, and F. Quevedo,
{\it Phys. Lett.} {\bf 187B} (1987) 25;
L. Ibanez, J. Kim, H. Nilles, and F. Quevedo,
{\it Phys. Lett.} {\bf 191B} (1987) 282;
L. Ibanez, J. Mas, H. Nilles,
and F. Quevedo, {\it Nucl. Phys.} {\bf B301} (1988) 157.}

\lref\binetruy{P. Binetruy and P. Ramond, \pl{350} (1995) 49,
and references therein.}

\lref\bachas{C. Bachas, C. Fabre and T. Yanagida, hep-th/9510094.}

%\lref\nilles{H. P. Nilles and S. Steiberger, hep-th/9510009.}

\lref\vw{C. Vafa and E. Witten, hep-th/9507050.}

\lref\hls{J. Harvey, D. Lowe, and A. Strominger, hep-th/9507168.}

\lref\ralph{R. Blumenhagen and A. Wisskirchen, hep-th/9506104;
R. Blumenhagen, R. Schimmrigk, and A. Wisskirchen, hep-th/9510055.}

\lref\clifford{P. Berglund, C. Johnson, S. Kachru, and P. Zaugg,
hep-th/9509170.}

\lref\dsw{ M. Dine, N. Seiberg, and E. Witten,
\np{289} (1987) 589; J. Attick, L. Dixon, and A. Sen,
\np{292} (1987) 109; M. Dine, I. Ichinose,
and N. Seiberg, \np{293} (1987) 253;
M. Dine and C. Lee, \np{336} (1990) 317.}

\lref\anamaria{ A. Font, L. Ibanez, H. Nilles, and F. Quevedo,
\pl{210} (1988) 101; A. Font, L. Ibanez,
H. Nilles, and F. Quevedo, \np{307} (1988)
109; A. Font, L. Ibanez, and
F. Quevedo, \pl{228} (1989) 79.}

\lref\casas{J. Casas and C. Munoz, \pl{209}
(1988) 214; {\bf 214B} (1988) 63; {\bf 212B} (1988) 343;
\np{332} (1990) 189; J. Casas, F. Gomez, and C. Munoz,
\pl{292} (1992) 42.}

\lref\zthree{ A. Font, L. Ibanez, and F. Quevedo,
\pl{217} (1989) 272; A. Font, L. Ibanez, F. Quevedo, and A. Sierra,
\np{331} (1990) 421.}

\lref\klt{ H. Kawai, D. Lewellen, and S.-H. H. Tye,
\np{288} (1987) 1; I. Antoniadis, C. Bachas, and C. Kounnas,
\np{289} (1987) 87.}

\lref\ten{H. Kawai, D. Lewellen, and S.-H. H. Tye,
\prd{34} (1986) 3794.}

\lref\nahe{ I. Antoniadis, J. Ellis, J. Hagelin, and D.
Nanopoulos, \pl{231} (1989) 65.}

\lref\finnell{ D. Finnell, hep-th/9508000.}

\lref\cchl{ S. Chaudhuri, S.-w. Chung, G. Hockney and J. Lykken,
hep-ph/9501361, to appear in {\it Nucl. Phys.}\/ {\bf B}.}

\lref\chl{ S. Chaudhuri, G. Hockney and J. Lykken,
{\it Phys. Rev. Lett.}\/ {\bf 75} (1995) 2274.}

\lref\cp{ S. Chaudhuri and J. Polchinski, hep-th/9506048.}

\lref\gauge{L. Ibanez and D. Lust, \pl{272} (1991) 251;
M. Gaillard and R. Xiu, \pl{296} (1992) 71.}

\lref\textures{J. Ellis and M. Gaillard, \pl{88} (1979) 315.}

\lref\strom{E. Witten, \np{268} (1986) 79; A. Strominger, \np{274}
(1986) 253.}

\lref\gins{P. Ginsparg, \pl{197} (1987) 139.}

\lref\faraggi{A. Faraggi, D. Nanopoulos and K. Yuan,
\np{335} (1990) 347;
A. Faraggi, \pl{274} (1992) 47; \pl{278} (1992) 131;
\np{387} (1992) 239.}

\lref\lnyb{J. Lopez, D. Nanopoulos and K. Yuan, \prd{50} (1994) 4060.}

%\lref\lnya{J. Lopez, D. Nanopoulos and K. Yuan, \np{399} (1993) 654.}

\lref\dreiner{H. Dreiner, J. Lopez, D. Nanopoulos and D. Reiss,
\np{320} (1989) 401.}

\lref\nelson{M. Dine, A. Nelson, and Y. Shirman, {\it Low energy
dynamical supersymmetry breaking
simplified}, Santa Cruz preprint SCIPP 94/21 (1994).}

\lref\sw{N. Seiberg and E. Witten, \np{276} (1986) 272.}

\lref\aldazabal{G. Aldazabal, A. Font, L. Ibanez, A. Uranga,
\np{452} (1995) 3; hep-th/9508033.}

\lref\abk{I. Antoniadis, C. P. Bachas, and C. Kounnas,
\np{289} (1987) 87; I. Antoniadis and C. P. Bachas, \np{298} (1988) 586.}

\lref\klst{H. Kawai, D. C. Lewellen, J. A. Schwartz, and S.-H. H. Tye,
\np{299} (1988) 431.}

\lref\davidl{D. C. Lewellen, \np{337} (1990) 61.}

\lref\font{A. Font, L. Ibanez, and F. Quevedo, \np{345}
(1990) 389.}

\lref\ffont{A. Font, L. Ibanez, F. Quevedo, and A. Sierra,
\np{331} (1990) 421.}

\lref\kap{V. Kaplunovsky, \np{307} (1988) 145; \np{382} (1992) 436.}

\lref\sgen{W. Lerche, A. Schellekens and N. Warner, Phys. Rep. 177 (1989) 1.}

\lref\sche{A. Schellekens, \pl{237} (1990) 363.}

\lref\bdfm{T. Banks, L. Dixon, D. Friedan and E. Martinec,
\np{299} (1988) 613.}

\lref\ds{M. Dine and N. Seiberg, \pl{162} (1985) 65; \np{301} (1988) 357.}

\lref\bd{T. Banks and L. Dixon, \np{307} (1988) 93.}

\lref\nao{S. Hamidi and C. Vafa, \np{279} (1987) 465;  D. Freed and
C. Vafa, \cmp{110} 349 (1987).}

\lref\chl{S. Chaudhuri, G. Hockney and J. Lykken, {\it Phys.
Rev. Lett.}{\bf 75} (1995) 2264.}

\lref\aso{K. Narain, M. Sarmadi and C. Vafa, \np{288} (1987) 551.}

\lref\dhvw{L. Dixon, J. Harvey, C. Vafa, and E. Witten, \np{261} (1985)
678; \np{274} (1986) 285.}

\lref\gep{D. Gepner, \np{287} (1987) 111; {\it ibid.}, \np{296}
(1987) 757.}

\lref\narain{K. S. Narain, \pl{169} (1986) 41; K. S. Narain,
M. H. Sarmadi and E. Witten, \np{279} (1987) 369.}

\lref\cleaver{G. Cleaver, {\it Supersymmetries in free fermionic
strings}, hep-th/9505080.}

\lref\keith{K. Dienes and A. Faraggi, {\it Phys. Rev. Lett.}
{\bf 75} (1995) 2646; hep-th/9505046; K. Dienes, A. Faraggi, and J.
March-Russell, hep-th/9510223.}

\lref\gross{D. Gross and J. Sloan, \np{291} (1987) 41.}

\lref\lust{D. Lust, S. Theisen, and G. Zoupanos, \np{296} (1988) 800.}

\lref\sunny{S. Kalara, J. Lopez, and D. Nanopoulos, \np{353} (1991) 650.}

\lref\search{J. Lopez, D. Nanopoulos, and K. Yuan, \np{399} (1993) 654.}

\lref\usgut{S. Chaudhuri, G. Hockney, and J. Lykken, to appear.}

\lref\iban{L. Ibanez, \pl{318} (1993) 73.}

\lref\niland{H.P. Nilles and S. Stieberger, hep-th/9510009.}

\lref\campbell{D. Bailey, B. Campbell, and S. Davidson,
\prd{43} (1991) 2314.}

\lref\lang{P. Langacker, talk at conference ``Unification: From
the Weak Scale to the Planck Scale'', ITP Santa Barbara,
Oct. 23-27, 1995.}

\lref\newfar{A. Faraggi, \pl{339} (1994) 223.}

\lref\nsv{K. Narain, M. Sarmadi, and C. Vafa, \np{288} (1987) 551;
\np{356} (1991) 163.}

\lref\schell{A. Schellekens, \pl{237} (1990) 363.}

\lref\nahethree{A. Faraggi and D. Nanopoulos, \prd{48} (1993) 3288;
A. Faraggi, \pl{326} (1994) 62.}

\lref\idr{L. Ibanez, D. Lust, and G. Ross, \pl{272} (1991) 251.}
%%%%%%%%%%%%%%%%%%%%%%%%%%%%%%%%%%%%%%%%%%%%%%%
\Title{\vbox{\baselineskip12pt\hbox{FERMILAB-PUB-95/349-T}
\hbox{NSF-ITP-95-134}\hbox{hep-th/9510241}}}
{\vbox{\centerline{
            \bf THREE GENERATIONS IN THE FERMIONIC CONSTRUCTION}}}
%\vskip2pt\centerline{\bf }}}
\centerline{{\bf S. Chaudhuri\footnote{$^\dagger$}
{e-mail : sc@itp.ucsb.edu}, G. Hockney$^*$ and J. Lykken\footnote{$^*$}
{e-mail : hockney@fnalv.fnal.gov, lykken@fnalv.fnal.gov}}}
%}
\vskip.1in
\centerline{$^{\dagger}$Institute for Theoretical Physics,}
\centerline{University of California, Santa Barbara, CA 93106}
\vskip.05in
%\centerline{$^{*}$Theory Dept., MS106,}
\centerline{$^{*}$Fermi National Accelerator Laboratory,}
\centerline{P.O. Box 500, Batavia, IL 60510}
\vskip .5cm
\noindent
We obtain three generation
$SU(3)_c$$\times$$SU(2)_L$$\times$$U(1)_Y$
string models in all of the exactly solvable $(0,2)$ constructions
sampled by fermionization.  None of these examples, including
those that are symmetric abelian orbifolds, rely on the
$Z_2$$\times$$Z_2$ orbifold underlying the NAHE basis.
We present the first known three generation models for
which the hypercharge normalization, $k_1$, takes values
smaller than that obtained from an $SU(5)$ embedding,
thus lowering the effective gauge coupling
unification scale. All of the models contain fractional
electrically charged and vectorlike exotic matter that could
survive in the light spectrum.

\Date{10/95}

\newsec{Introduction}

Recent developments in superstring model building have
focused on constructions with $(N_L,N_R)$$=$$(0,2)$ world-sheet
supersymmetry \refs{\aldazabal,\cchl,\kachru,\ralph,\clifford}.
This is in part due to the difficulty in obtaining low numbers
of generation-anti-generation pairs in the simpler class
of three generation $(2,2)$ compactifications
\refs{\cyphen,\rolf,\gepner}. Despite these technical advances,
the sample of three generation models available
in sufficient detail to enable phenomenological
analyses have remained the examples of symmetric $(0,2)$ orbifold
models, obtained in the $Z_3$ orbifold
\refs{\fernando,\anamaria,\zthree,\casas}
and free fermionic constructions \refs{\nahe,\faraggi}.

There are new developments in understanding
strongly coupled string theory which could lead to a different
formulation better suited to exploring low energy string theory.
These developments have thus far had little
direct impact on our understanding of $N$$=$$1$ string theories
but this may change. We caution the reader at the outset
that it is extremely unlikely that every feature of the tree
superpotential and massless spectrum of any {\it particular}
classical $N$$=$$1$ heterotic vacuum survives quantum corrections.
Thus, the objective of such phenomenological analyses is not to
arrive by accident at a fully realistic model but rather to sample
vacua for {\it generic} features that would be unanticipated
in traditional unified field theories.

In this paper, we resolve the problem of obtaining {\it odd}
generation number, with generic gauge group and generic
matter content, in the fermionic construction
\refs{\ten,\klt,\klst}. We will use the new formalism
for real fermions developed in \refs{\cchl}.
It is well-known that the free fermionic models
\refs{\klt,\abk} are equivalent to abelian orbifold models,
with symmetric $Z_2$$\times$$Z_2$ point group twists and
quantized $Z_N$ Wilson lines. The real fermion construction, on
the other hand, {\it samples} the full range of exactly solvable
$(0,2)$ constructions. This includes the more generic class
of asymmetric and nonabelian orbifolds \refs{\nsv}. It also
includes an asymmetric generalization of the Gepner
construction: compactifications based on tensor products
of $18$ right and $44$ left-moving $c$$<$$1$ conformal field
theory building blocks.

As a beginning, we provide fermionic realizations of seven new
exact conformal field theory (cft) solutions embedding
three generations of chiral superfields transforming under
$SU(3)_c$$\times$$SU(2)_L$$\times$$U(1)_Y$, realized at
Kac-Moody level one, along with the Higgs multiplets of the
minimal supersymmetric standard model (MSSM).
All of these solutions contain an anomalous gauged
$U(1)$ flavor group at tree level.
We can analyse the tree superpotential for all possible flat
directions along which this anomalous $U(1)$ can be broken, giving
families of nearby supersymmetric ground states \refs{\dsw}.
The solutions also contain nonanomalous gauged $U(1)$ flavor groups.
We find considerable freedom in the separation of
the hypercharge embedding from the nonanomalous $U(1)$'s.
A consistent vacuum shift to remove the anomaly and a
consistent embedding of hypercharge defines a string model.
We can then compute the effective superpotential to arbitrarily
high order in the nonrenormalizable contact terms of these models,
thus making phenomenological analysis possible.

The examples in this paper contradict some of the folklore of
fermionic string model building. There is no indication of the so-called
\lq\lq unique" NAHE basis in the fermionization of any of our three
generation examples, {\it including} the three
examples based on free fermions,
i.e., Ising and Weyl fermions, alone. It has also been speculated in
the literature that no fermionic constructions exist of heterotic
string models based on embeddings of the group weights and hypercharge
assignments of the quarks and leptons of the MSSM with
$k_1$$<$${{5}\over{3}}$ \keith . We find several explicit
counter-examples. The absence of a lower bound on $k_1$ appears in fact
to be quite generic and is independent of the fermionic construction.

\newsec{Review of Construction}

To obtain genuinely distinct three generation models in the fermionic
construction with generic gauge groups and generic matter content,
we have introduced several new features in the underlying fermionic
representation theory. Modular invariance of the one-loop vacuum
amplitude and associativity of the vertex operator algebra restricts
fermionic realizations of conformal field theory solutions to
string theory as follows. The individual
Majorana-Weyl world-sheet fermions must be
paired into one of three possible fermionic cfts: Weyl fermions, Ising
fermions, or in blocks of chiral
Ising fermions. The consistent choices of spin structure for blocks
of chiral Ising fermions allowed by associativity of the fermionic
cfts were analyzed in \cchl .

\subsec{Holomorphic Rank Reduction}

If all of the chiral Ising fermions in a block are left-moving, this
corresponds to a holomorphic cft of central charge $c_{m}$$=$$8$, $12$,
$14$, $15$, $16$, $\ldots$ $22$ \cchl .
Such holomorphic cfts give rise to
rank reduction in the string model. The earliest known example is
the tachyonic ten-dimensional modular invariant discovered in
\refs{\ten} with $248$ gauge bosons, i.e., a single $E_8$ at level two.
Despite their intriguing properties rank reduced models have been
difficult to work with until recently because of the lack of a
straightforward prescription for identifying physical states and
superpotential couplings given the one-loop partition function.

Such holomorphic cfts can be tensored together
with chiral boson cfts to
build $N$$=$$4$ supersymmetric models which do not correspond to
Narain compactifications. At generic points in the moduli space
the gauge group is $(U(1))^{28-c_m}$ and the dimension of the
moduli space is given by $r_L r_R$, where $(r_L,r_R)$$=$$(22-c_m,6)$.
The $N$$=$$4$ fermionic models with rank reduction $c_m$$=$$8$, $12$,
and $14$ have been interpreted as asymmetric orbifolds \refs{\cp}.
The simplest example is a $Z_2$ orbifold of the toroidally
compactified $E_8$$\times$$E_8$ string, where one mods out by the
outer isomorphism which interchanges the two $E_8$ lattices accompanied
by a shift of half-periodicity in any cycle of the torus.
All three rank reductions can be obtained by compactifying the
$SO(32)$ string on an $(SO(4))^3$ torus and introducing
the following Wilson lines on the gauge lattice
\eqn\wil{
\eqalign{
&(1,1,1,1,1,1,1,1,0,0,0,0,0,0,0,0)\cr
&(1,1,1,1,0,0,0,0,1,1,1,1,0,0,0,0)\cr
&(1,1,0,0,1,1,0,0,1,1,0,0,1,1,0,0)\cr
}}
which breaks the gauge symmetry to
\eqn\lattice{
\eqalign{
 \big[ (SO(4)\times & SO(4)) \times ((SO(4)\times SO(4)) \big] \times \cr
 \big[ &(SO(4)\times SO(4) )\times (SO(4)\times SO(4)) \big]
\times (SO(4))^3
}}
This gauge lattice has three commuting $Z_2$ isometries
under interchange of repeated $(D_2)^n$ units, $n$$=$$1$,$2$,$4$,
together with their conjugacy classes. Modding out by the
$Z_2$ isomorphisms sequentially accompanied by shifts of
half-periodicity in the $(SO(4))^3$ torus gives models with
enhanced symmetry points at which part of the gauge group is realized
at Kac-Moody level $2$, $4$, and $8$, respectively.

Combining such $Z_2$ twists on the left-moving gauge lattice
with twists on the right-moving world-sheet fermions gives $N$$=$$1$
asymmetric orbifolds of reduced rank. There is an interesting
distinction between this mechanism for achieving higher level, first
used in the fermionic construction by Lewellen \refs{\davidl}, and
that employed in the higher level orbifold models of \refs{\font,\aldazabal}.
(Note that only the asymmetric mechanism can apply in the $N$$=$$4$
case.) Unlike the symmetric orbifolds, the point in the moduli
space with level two gauge group $G$ is {\it infinitely} distant from the
level one $G$$\times$$G$ point and corresponds to a decompactification
limit \cp . This is the large radius limit of the circle in which we
embed the accompanying shift.

Holomorphic rank reduction should not be confused with
field theoretic Higgsing. In the latter case, the rank of the
gauge group varies {\it locally} within the moduli space as one varies
the vev of a Higgs field transforming in the fundamental
representation of the gauge group.

The case of $N$$=$$1$ supersymmetry allows a much richer set of
possibilities than the $N$$=$$4$ case, giving examples both of asymmetric
orbifold and heterotic Gepner model constructions.
In an $N$$=$$1$ solution, the block of chiral Ising fermions
can be split among $n_L$ left-moving and $n_R$ right-moving fermions, such
that $n_L$$+$$n_R$$=$$c_m$ takes one of the allowed values listed
above \cchl . For the purposes of this paper, we consider fermionic
realizations where all of the right-moving world-sheet fermions are
Majorana-Weyl, with periodic or anti-periodic boundary conditions alone.
This class of fermionic solutions already includes several
new possibilities
for three generation models.

Other models have a fermionic realization that includes {\it right-moving}
chiral Ising fermions. They are the first known examples of exact cft
solutions to string theory based on holomorphic tensor products of
$c$$<$$1$ building blocks. A bosonic description of the underlying
target space of such solutions is at this point unknown.
We note that in abstract $(0,2)$ cft constructions obtaining
a one-loop modular invariant partition function does {\it not} by itself
provide
enough information about a solution to string theory. It is essential to
develop
a formalism (i) that unambiguously identifies (physical) states in
the partition function with string vertex operators,
and (ii) constructs the
vertex operator algebra yielding the complete tree superpotential including
nonrenormalizable terms and couplings to singlets. The nonrenormalizable terms
and the singlet couplings are crucial since they
probe the moduli space of flat directions in the neighbourhood of the
exactly solvable point.

The formalism used here and
developed in \cchl\ has a natural extension
to heterotic tensor products of other holomorphic cfts with $c$$<$$1$.

\subsec{Overlapping Embeddings}

To enlarge the scope of free
fermionic representation theory we make some further modifications.
We allow overlapping embeddings of the current algebra weights
into fermionic charges, $Q_F^i$,
where $i$ labels individual Weyl fermions, and $G$ and $G'$
are given commuting current algebras,
\eqn\overlap{
w_{G}^i + w_{G'}^i ~=~ Q_F^i
}
Thus, in many of our conformal field theory solutions the group
weights of the hidden and the visible gauge groups actually
overlap! This has no bearing on spacetime physics or equivalently
on the conformal field theory, but is simply a trick that allows a
free fermionic representation for many new modular invariant
partition functions. Enlarging the class of allowed embeddings
considerably reduces the ad-hoc restrictions on groups/weights
obtained in conventional free fermionic solutions \cchl .
This increased flexibility in choosing the fermionic embedding of
the gauge group is crucial in obtaining three generations
{\it simultaneously} with generic gauge group and generic matter
content.

As an example, consider the following embedding of the simple
roots $(\beta_i , \delta ; ~ \alpha_j  )$, of
$SU(3)_c$$\times$$SU(2)_L$$\times$$SU(4)_2$, which appears in Model 4.
The $SU(4)_2$ plays the role of a confining
hidden sector group in this model. The roots are embedded in eleven
fermionic charges as follows:
\eqn\embed{
\eqalign{
\beta_1 &=  (\ha ,\ha ,\mha ,\mha ,\mone ,\ze ,\ze ,\ze ,\ze ,\ze ,\ze ) \cr
\beta_2 &=  (\ze ,\ze ,\ze ,\one ,\ha ,\ha ,\mha ,\ha ,\ze ,\ze ,\ze ) \cr
\delta &= (\ha ,\ha ,\mha ,\mha ,\one ,\ze ,\ze ,\ze ,\ze ,\ze ,\ze ) \cr
\alpha_1 &= (\ze ,\ze ,\ze ,\ze ,\ze ,\ha ,\ze ,\mha ,\ha ,\ze ,\mha ) \cr
\alpha_2 &= (\qt ,\qt ,\qt ,\qt ,\ze ,\mha ,\ze ,\ze ,\mha ,\ha ,\ze ) \cr
\alpha_3 &= (\ze ,\ze ,\ze ,\ze ,\ze ,\ha ,\ha ,\ze ,\ze ,\mha ,\ha ) \cr
}}
There are five overlapping, but orthogonal, $U(1)$ generators,
spanning the remainder of the embedding space.

\subsec{Twist Field Current Algebra Realizations}

In addition to conventional fermion bilinear currents we also
consider twist field realizations for part, or all, of the
current algebra. Such currents are obtained by tensoring
together $8$ or $16$ dimension ${{1}\over{16}}$ twist
operators of the block of chiral Ising cfts (tensored also with
a dimension $\half$ fermion operator in the first case),
so as to give holomorphic
operators of dimension $(1,0)$.
There is also a similar construction with $4$ or $8$
dimension ${{1}\over{8}}$ Weyl twist fields.
For example,
\eqn\twist{
J_{ijkl}(z) ~=~ j_{free} (z)
\left ( \sigma_i^+ \sigma_j^+ \sigma_k^+ \sigma_l^+ ~+~
\sigma_i^- \sigma_j^- \sigma_k^- \sigma_l^- \right )
}
where $i$$\neq$$j$$\neq$$k$$\neq$$l$ label four inequivalent
pairs of fermions in the chiral Ising block, and $j_{free}$ is
the product of four dimension ${{1}\over{8}}$ twist fields in the free
field (Weyl fermion) cft. Current algebra realizations
combining twisted currents with conventional fermion bilinear
currents abound in the fermionic construction. They play an
essential role in fermionic realizations of higher level
and non-simply laced gauge symmetry, as in the examples
of \chl . But they also provide new fermionic realizations
of the level one simply laced gauge groups.

One can also find examples in which all of the currents
including Cartan generators have a twist field
realization! In such a model, none of the gauge bosons
would appear in the untwisted sector i.e., in the sector
in which all of the world-sheet fermions obey NS boundary
conditions. Twisted current algebra realizations are a
source for {\it accidental} extensions of the gauge symmetry,
originating in the block of chiral Ising fermions.

\subsec{World-sheet supersymmetry}

The examples in this paper were constructed using
the conventional fermion trilinear
generator of the $(1,0)$ world-sheet
supersymmetry of the heterotic string
\eqn\supercurrent{
T_F ({\bar{z}}) ~=~ i \sum_{\mu = 1}^{2} \psi_{\mu}
\partial_{{\bar{z}}} X^{\mu} ~+~ i \sum_{k=1}^{6}
\psi_{3k} \psi_{3k+1} \psi_{3k+2}
}
where the index $\mu$$=$$1,2$ sums over the two transverse
dimensions in $D$$=$$4$, and we work in light-cone gauge \klt .
The $N$$=$$1$ spacetime supersymmetry charges are embedded in the
spin structure of eight right-moving fermions, which are
paired into four Weyl fermions as follows, $\psi_1 + i \psi_2$,
$\psi_{3k} + i \psi_{3k+3}$, $k$$=$$1,3,5$. The remaining $12$
right-movers can be Weyl, Ising, or chiral Ising fermions.
The left-moving Weyl fermions are unrestricted
by world-sheet supersymmetry and are allowed to satisfy
any rational boundary condition. Further generalization to
rational boundary conditions on the right moving Weyl fermions
is possible. Such right moving spin structures have been analyzed
recently in \refs{\cleaver}, but have not as yet been incorporated
into actual models.

\subsec{Couplings}

Given a conformal field theory solution we can compute
arbitrary N-point functions of the vertex operators
which represent massless physical states.
{}From these string tree level S-matrix elements
we can then deduce the effective field theory
action in a derivative expansion \refs{\gross}.
The effective superpotential thus derived contains
no quadratic terms, but does generally contain
cubic terms as well as quartic and higher order
nonrenormalizable contact terms.

To compute the effective superpotential it suffices
to consider N-point cft correlators of the
form \refs{\lust ,\sunny}

\eqn\corr{
\eqalign{
\int\prod_{i=1}^{N-3}\, d^2z_i\, \Big<
&V_{1(-1/2)}^f(z_{\infty})\, V_{2(-1/2)}^f(1)\,
V_{3(-1)}^b(z_1) \cr
&V_{4(0)}^b(z_2)\ldots V_{N-1(0)}^b(z_{N-3})\,
V_{N(0)}^b(0) \Big> \cr
}}
where $V_{(-1/2)}^f(z_i)$ is the vertex operator for the
fermionic component of a chiral superfield in the
ghost number $-1/2$ picture, while
$V_{(-1)}^b(z_i)$, $V_{(0)}^b(z_i)$ are vertex operators
for the bosonic components of chiral superfields in the
ghost number --1 and 0 pictures, respectively. $SL(2,C)$
invariance was used, as usual, to fix three of the $z_i$
to $\infty$, 1, and 0.

For the Weyl fermion and ghost cfts, the computation of
N-point correlators is straightforward. For the Ising fermion
cfts, the computations are more involved but are available
in the literature \sunny .

The N-point correlators of the cfts described by blocks
of chiral Ising fermions require more work. In \refs{\cchl}
we derived the selection rules for which chiral Ising
correlators can be nonzero. These selection rules
can be traced to spin one simple currents in the
conformal field theory. These holomorphic dimension (1,0)
operators are not true conserved
currents, because they are not local with respect to
at least one of the physical vertex operators which does
{\it not} appear in the specified correlator.
Nevertheless the $n$ simple charges thus defined are
conserved in the correlator, otherwise the correlator vanishes.

The computer implementation of the selection rules
arising from the chiral conformal field theories
is still incomplete, so we caution the reader that
some of the terms listed in the superpotential
given in Tables 4.1-4.3 may eventually vanish.
At the moment, we are limited to checking
this by hand.

\newsec{General Aspects of Model Building}

It is helpful to keep in mind a number of phenomenological issues
when building possibly semi-realistic string models. We summarize
these below.

\subsec{Gauge Embedding}

A perturbative ground state of heterotic string
theory (a string model)
provides an effective field theory description of
physics at the string scale:
$M_{str} \simeq g_{str}$$\times$$5$$\times$$10^{17}$ GeV.
A starting point for obtaining string models is to
find an exact conformal field theory solution to
heterotic string theory.
In solutions with $N$$=$$1$ spacetime
supersymmetry, the rank of the full gauge group
is $\le 22$.
It may be possible
to find solutions
such that the full gauge group at the string scale is precisely
$SU(3)_c$$\times$$SU(2)_L$$\times$$U(1)_Y$ of the
MSSM, however no one has yet succeeded
in constructing such a solution. In existing solutions the full
gauge group is of the form:
\eqn\fullgauge{
G_{SM} \times G_{flavor} \times G_{hidden}
}
Here $G_{SM}$ is either the standard model gauge group or a larger
nonsimple group which embeds it, $G_{flavor}$ represents new
gauge interactions (typically a product of $U(1)$'s) under which
the quarks and leptons of the MSSM transform in a flavor-dependent
way, and $G_{hidden}$ represents new gauge interactions of particles
in a hidden (or semi-hidden) sector.

Since gauged flavor symmetries and hidden sectors are useful
for inducing fermion mass hierarchies and dynamical supersymmetry
breaking, respectively, there is no obvious necessity in building
semi-realistic string models to reduce the rank of the cft solution
below $22$. However, the generic class of solutions does in fact
include such rank reduction.

In some previously known solutions $G_{SM}$ is just
$SU(3)_c$$\times$$SU(2)_L$$\times$$U(1)_Y$.
In these solutions,
the conformal current algebra which realizes this gauge algebra
has Kac-Moody levels $k_3 = k_2 = 1$. This has the beneficial
effect of restricting the massless chiral supermultiplets
in such solutions to
be triplets, antitriplets, or singlets under $SU(3)_c$, and
doublets or singlets under $SU(2)_L$. However, higher level
embeddings of $G_{SM}$ are the more generic
class.\foot{Note that level one also necessitates fractionally
charged states in the string spectrum (see section 3.5). Higher
level removes this restriction.}

Other solutions,
such as flipped $SU(5)$, take $G_{SM}$ to be a larger nonsimple
or nonsemisimple group that embeds the standard model group.
The known solutions of this type also have Kac-Moody level equal
to one for the nonabelian factors of $G_{SM}$. None of these
solutions is a good starting point for constructing a
conventional GUT,
and they cannot be since Kac-Moody level
one excludes the possibility of massless adjoint Higgs at the
string scale. Such solutions require Kac-Moody level two or greater,
which in turn requires rank reduction to embed the higher-level
gauge group. It should be possible to construct semi-realistic
higher level solutions,
but at the moment the best example is a three generation
SU(5) level two model which
suffers from light color sextet exotics \refs{\usgut}.

\subsec{Anomalous U(1)}

Many tree level cft solutions to string theory,
including those discussed here,
contain a U(1) gauge
factor which is anomalous. When this occurs, the Green-Schwarz
mechanism breaks the anomalous $U(1)$, at the expense
of generating  a Fayet-Iliopoulos $D$ term proportional to
\eqn\dflat{
D_A ~=~ \sum_i Q_i^A |\chi_i|^2 + {{g^2}\over{192 {\pi}^2}} e^{\phi}
{}~ Tr Q^A ~~~~,
}
where $\phi$ is the dilaton, $TrQ^A$ is the trace anomaly, and the
$\chi_i$ are scalar fields with anomalous charge $Q_i^A$.
This term will break supersymmetry and destabilize
the vacuum \refs{\dsw}.
The vacuum becomes stable and supersymmetry is restored when one
or more of the scalar fields which carry nonzero anomalous charge
acquire a vev such that the right-hand side of \dflat\ vanishes.
Supersymmetry is then restored provided that this vacuum shift
is in a direction which is F-flat and also D-flat with respect
to all non-anomalous $U(1)$'s. If we let $\chi_i$ now denote the
scalar {\it vevs} which cause \dflat\ to vanish, then the additional
D and F flatness constraints are
\eqn\fflat{
D_a ~= ~ \sum_i Q_i^a | \chi_i |^2 ~=~0 ,
{}~~<{{\partial W}\over{\partial \phi_j}}> ~=~0 ~~~~.
}
where $a$ labels the nonanomalous $U(1)$'s, and the $\phi_j$ are
all the chiral superfields, not just those whose scalar components
get vevs.

Note that the shifted vacuum is no longer a classical string vacuum,
but does correspond to a consistent perturbative quantum string vacuum.
Thus cft solutions which contain an anomalous $U(1)$ in some sense
access a much larger class of perturbative string vacua than those
that do not.

Note also that because the D term cancellation in
\dflat\ involves the one-loop generated anomaly, the scale of vevs
in \dflat\ is naturally (depending on the value of the anomaly)
smaller than the string scale, by an order of magnitude or so.
Since the scalars whose vevs $\chi_i$ contribute to \dflat\
often carry a variety of other abelian and nonabelian quantum numbers,
the vacuum shift generically breaks the original gauge group
to one of smaller rank. This rank reduction is variable and can
be quite large. It may be possible to perform this vacuum shift
without breaking the standard model gauge group, although there
is no fundamental reason why this should {\it always} be the case.
In fact, in many of our solutions we have found considerable
freedom in choosing the flat directions involved in the shift.
In some, but not all vacua, it is possible to break
$G_{flavor}$ completely at this stage \refs{\casas,\nahe,\faraggi}.

After the vacuum shift a number of previously massless fields
will acquire masses, of order $(\alpha_{str})^nM_{str}$
for some $n$, via coupling
to scalar vevs. The spectrum of light fields, particularly light
exotics, is often much reduced. In addition, the scalar vevs also
tend to induce a number of effective Yukawa interactions for the
MSSM quarks and leptons, with Yukawa couplings that are naturally
suppressed by powers of $\alpha_{str}$.
This combination of favorable outcomes were employed first in
orbifold models \refs{\fernando,\anamaria,\zthree,\casas},
then in the free fermionic
construction\refs{\nahe,\faraggi}, to make a first pass at
viable perturbative superstring phenomenology.

\subsec{Three Generations}

One of the striking things about all known superstring contructions
is the difficulty of finding vacua with precisely three
chiral generations. Thus despite the plethora of perturbative superstring
vacua there is a paucity of three generation constructions, and this is
the main reason why the number of known semi-realistic models is still
so few.

This problem has been much discussed in the
free fermionic construction \refs{\nahethree}.
All of the previously known three generation solutions
\refs{\nahe,\faraggi,\finnell} are
based on a $Z_2$$\times$$Z_2$ orbifold of the heterotic string
compactified on an $SO(12)$ six-torus with arbitrary background fields.
Each successive $Z_2$ twist of this torus breaks half of the spacetime
supersymmetries, and the untwisted sector contains the moduli deformations
of an $(SO(4))^3$ torus. This leads to the so-called NAHE basis \nahe\
for the basis vectors which specify the fermion spin structures.

Although the specifics of these solutions vary,
the fermionic construction allows very few distinct solutions
for a given gauge group {\it once} the NAHE embedding of three
generations of chiral matter fields has been imposed.
Although there has been considerable speculation that the
NAHE basis is necessary to obtain three generation solutions
in the free fermionic construction, this is not the case.

By changing the embedding of the
standard model gauge group and the chiral matter fields we
have produced a large number of three generation solutions
which have no connection to the NAHE basis.

\subsec{Gauge Coupling Unification}

A remarkable property of string theory is that
it provides gauge coupling unification independently of
grand unification into a simple group.
At string tree level:
\eqn\gunify{k_3g_3^2 = k_2g_2^2 = k_1g_1^2 = g_{str}^2
}
Here $k_1$ is not a Kac-Moody level, but rather a normalization
factor which relates the hypercharge $q$ of a state with the
hypercharge contribution to the conformal dimension $h$, of the
conformal field which creates the state:
\eqn\handg{h = {q^2\over 4k_1}
}
In a given cft solution the conformal dimensions of the fields
are fixed. Thus $k_1$ can be determined in any solution
by, for example, declaring the quark doublet states to have their
conventional hypercharge $q$$=$$1/3$, then using \handg\ to compute $k_1$.
Because the total conformal dimension of a
field which creates a massless string state must
be $\le 1$, we see from \handg\ that any cft solution which
contains the right-handed electron multiplet of the MSSM must have
$k_1 \ge 1$ \refs{\schell}.

String solutions which embed the standard model group
into $SU(5)$ or $SO(10)$ conformal algebras have $k_1$$=$$5/3$.
All previously known cft solutions have
$k_1$ larger than $5/3$ --- usually much larger \refs{\keith}.
Of course $k_3$$=$$k_2$$=$$1$, $k_1$$=$$5/3$ makes \gunify\
resemble the putative gauge coupling unification of the MSSM,
{\it under the assumption that the visible spectrum is
exactly that of the MSSM},
and at a scale $M_{str}$ which is roughly one order of
magnitude higher than the $M_U \simeq 3$$\times$$10^{16}$ GeV
suggested by low energy data. String threshold corrections
to \gunify\ \refs{\kap,\idr} may explain the mismatch,
although this is not the case in the simplest abelian
orbifolds and free fermionic
models in which these have been analyzed (see \refs{\keith,\niland}
and references within). Unfortunately, the moduli dependence of such
thresholds is poorly understood in semi-realistic models and clearly
deserves further analysis \refs{\niland}. We can also
achieve agreement between string unification and
MSSM gauge coupling unification by lowering the value of $k_1$
by about 10-15\% \refs{\iban }.
We have found the first three generation string models
for which $k_1$$<$$5/3$, including an example (our Model 4)
with $k_1 = 1.458$.

Another possible explanation
of the mismatch in gauge coupling
unification is that there is a
separate grand unification scale,
(either $SU(5)$ or $SO(10)$),
with $SU(5)$ broken at $M_U$. Although it
may be possible to construct a classical string vacuum that mimics
the spectrum and couplings of a semi-realistic grand unified
model (see for example \refs{\bachas,\finnell})
it is unlikely that one could
determine $M_U$ within string perturbation theory. Such
a scenario would of course inherit the usual difficulties of
grand unified models. But this is a possible option.

Lastly, it is always possible (even within some string
models \keith ) to arrange for suitable combinations of exotic
particles at suitable mass scales, and thus change the
renormalization group (RG) running of the gauge couplings to
remove the mismatch. Since exotics are present in all known
string models, this solution may be less unnatural
than it appears. However it is very difficult to implement this
scenario without arranging {\it large} but nearly cancelling
effects in the RG equations. Thus, if this is the solution to
the mismatch problem, then the fact that string tree-level unification
and MSSM gauge coupling unification agree as well as they do
must be regarded as an accident!

The known cft solutions all contain a fairly large
number of light exotics before the anomalous $U(1)$ vacuum shift.
These will wreak havoc with the gauge coupling RG equations unless
almost all of them (a) acquire superheavy masses by coupling to
string scale vevs or the vevs involved in the vacuum shift, or (b)
assembling into approximate $SU(5)$ multiplets, which have a much
smaller effect on the RG running. Our solutions have similar
features, but to see whether they fare better or worse than
previous models requires detailed analysis of the vacuum shift,
couplings, and RG running of the effective couplings including
SUSY-breaking effects.

\subsec{Fractional Charge}

All $SU(3)_c$$\times$$SU(2)_L$$\times$$U(1)_Y$
cft solutions with
$k_3$$=$$k_2$$=1$ must contain exotics with fractional
electric charge \schell . This is because,
if all physical states in a string vacuum obey charge quantization,
there exists a certain conformal operator which
is mutually local with respect to all the physical fields. This
operator must thus itself correspond to a physical field, which leads
to a contradiction unless $k_1$$=$$5/3$ and the standard model gauge
group is promoted to unbroken $SU(5)$ at level one.

This argument does not determine whether or not there are
any {\it massless} string states with fractional charge;
it may be possible to arrange for fractional charges to occur
only in the {\it massive} modes of the string, and thus be superheavy.
However in all of the known models, fractionally charged
exotics do occur at the massless level.
It may be possible to avoid fractionally charged exotics
entirely in three generation string models with higher Kac-Moody
levels, but this has never been demonstrated.

The lightest fractionally charged particle will be stable. This
can create conflicts with experimental bounds from direct searches,
as well as rather
severe cosmological and astrophysical bounds \refs{\campbell}.
For example, the lightest fractionally
charged particle will completely dominate the energy density in
the universe if its mass is greater than
a few hundred Gev\refs{\lang,\campbell}.
If there is an inflationary epoch and subsequent reheating, we
can probably tolerate a lightest fractionally charged particle
with mass greater than the reheating temperature.

In all known string models,
a variety of fractionally charged exotics
are seen to occur. They can be $SU(3)_c$$\times$$SU(2)_L$ singlets
with hypercharges less than $\pm 2$, and
they can be color triplets or Higgs
with nonstandard hypercharge. These exotics have important effects
on the RG running of the couplings.

\subsec{Hypercharge Embedding and Particle Identification}

In cft solutions
for which $G_{SM}$$\times$$G_{flavor}$ is
$SU(3)_c$$\times$$SU(2)_L$$\times$$U(1)^n$, for some $n$,
particle identification is not automatic,
since string theory does not label the physical states for us.
For example, the three
lepton doublets and the up and down type Higgs doublets all
have the same $SU(3)_c$$\times$$SU(2)_L$
quantum numbers. Most known solutions also contain
a number of additional weak doublet exotics.

Hypercharge disentangles these doublets somewhat, but
string theory does not label hypercharge for us either, i.e.
it does not tell us how to extract $U(1)_Y$ from the additional
nonanomalous $U(1)$'s which go into $G_{flavor}$. In the known
solutions there is usually more than one consistent embedding
of $U(1)_Y$ as a linear combination of the original nonanomalous
$U(1)$'s.

Because of this there are many different ways of embedding the
standard model particle content within the {\it same}
conformal field theory solution of heterotic string theory.
Different choices of hypercharge and particle identification
will lead to different couplings, masses, and mixings of the
MSSM quarks and leptons, as well as different hypercharge
assignments for the exotics.  In practice we thus
use phenomenological considerations as a guide for
making these choices.

A given hypercharge embedding fixes $k_1$.
We will constrain the hypercharge embedding by requiring
that $k_1$ be reasonably close to $5/3$, and that the number of
fractionally charged states is minimized. We discuss this
procedure in detail for two of our models in section 4.
The reason for this additional flexibility in the
embedding of hypercharge is related to the fact that the
solutions are {\it not} based on
Wilson line breakings of $SU(5)$ or $SO(10)$ but instead
explore the generic class of gauge embeddings.

\subsec{Rapid proton decay}

String models typically violate matter parity,
allowing for the appearance of B and L violating terms in
the cubic part of the effective superpotential. In
particular, terms of the form
\eqn\badterms{Q\,L\,d^c + u^c\,d^c\,d^c
}
where $Q$ denotes a quark doublet, $L$ a lepton doublet,
and $u^c$, $d^c$ the conjugates of the right-handed up and down quarks,
would lead to instantaneous proton decay. In addition to these cubic
terms, there is also
the possibility of quartic terms which can lead to unacceptably
rapid proton decay.

To check a particular string model for the absence of
such dangerous terms, it is insufficient to
compute the effective superpotential to quartic order.
This is because the dangerous B violating terms may
be generated at {\it any} order via nonrenormalizable terms
which are unsuppressed due to string scale vevs.
The simplest solution to this problem
gauged $U(1)_{B-L}$ as part of $G_{flavor}$ \refs{\font}.
Other possibilities that have been considered in the
known models are a combination of B-L and custodial $SU(2)$
along with other flavor symmetries
which distinguish quarks from leptons \refs{\faraggi,\newfar}.

\subsec{Quark and Lepton Masses}

A major challenge for any unified model is to
reproduce, even qualitatively, the many observed hierarchies
of masses and mixings for quarks and leptons. In known cft solutions
the numerical values of the couplings in the effective superpotential
are order one, and this is likely to be true rather generally
in perturbative string vacua. Thus,
small Yukawa couplings in the MSSM may originate
from scalar vevs or fermion condensates which
take values at scales other than $M_{str}$. Nonrenormalizable
couplings of quarks and leptons to these vevs or condensates
can then generate effective Yukawa couplings which are small.

A beautiful property of the known string models
is that such a mechanism does indeed occur: the vacuum
shift associated with the anomalous $U(1)$.
It appears unlikely that any one mechanism will
explain all of the observed hierarchies.
Some previously known models \refs{\nahe,\faraggi}
can produce a top quark Yukawa which is order one, while
all the other effective Yukawas are suppressed by an
order of magnitude or more. We show below that least
one of our models shares this feature.

\newsec{Models}

To illustrate the range of options within the
real fermion construction for obtaining precisely
three generations of chiral matter, we have constructed
a sample of
seven new
conformal field theory solutions embedding the standard model
gauge group.
Table 1 lists for each solution $c_m$, where $22$$-$$c_m$ is
the rank of the
full gauge group,
the number, $n$, of nonanomalous $U(1)$'s in
$G_{flavor}$, the number of vectorlike pairs of
color triplet exotics,
the number of extra weak doublets,
$G_{hidden}$, and whether or not there is chiral matter
transforming under $G_{hidden}$.
It should be noted that these
are properties of the conformal field theory solutions
{\it before} making the vacuum shift required by the
presence of an anomalous $U(1)$ and the choice of $k_1$
which together define a string model. In these solutions
$G_{flavor}$ always contains an anomalous
$U(1)$ and at least 5 additional $U(1)$'s.
As can be seen for example from
Tables 2.1 through 2.3,
the quarks and leptons carry
complicated, highly flavor dependent charge assignments under these
extra $U(1)$'s. Some or all of these extra $U(1)$'s will be broken
by the vacuum shift.
It is also clear that
there are particles which are both nonsinglets
under $G_{hidden}$ and carry hypercharge, charges under $G_{flavor}$,
or are weak doublets. However after the vacuum shift
many of these particles
will become superheavy, and the rank of $G_{flavor}$ (and perhaps
$G_{hidden}$) is reduced. It is possible
that after the vacuum
shift $G_{hidden}$ is truly hidden.
Note that $G_{hidden}$ need not be simple or semisimple,
does not necessarily
have any large nonabelian factors, and can have higher
Kac-Moody levels. There is also typically chiral matter
in the hidden sector. All of these features may be important
for scenarios of dynamical supersymmetry breaking.

{}From Table 1 we see that the number of exotics in these models
varies considerably from model to model. After the vacuum shift
some of these exotics become superheavy, while the others will
acquire TeV or intermediate scale masses after supersymmetry
breaking.

Many features just described are similar to
previously known solutions\fernando\anamaria\casas\faraggi . Let us
now focus on the features of these solutions
which are qualitatively new.

\subsec{Three generations}

The previously known three generation solutions
in the free fermionic formulation all obtain three chiral
generations by using the NAHE basis. In this construction
the three generations arise from three distinct sectors with
different left and right-moving spin structure, corresponding
to the three twisted sectors of a symmetric $Z_2$$\times$$Z_2$
orbifold. In our solutions the three generations arise from
sectors (i.e.
choices of fermionic spin structure) in a variety
of new ways. This can be seen from Table 3, which for each solution
lists the sectors of the cft which contains the
positive helicity fermionic component highest weight
states of the three generations of chiral superfields.
The sectors are listed as linear
combinations of the basis vectors for each fermionic cft solution;
the basis vectors
are given in Tables A.1 to A.7 of the
appendix. From these tables we can summarize the different
realizations of three generations as follows:

\noindent{\it Free fermionic realizations:} these solutions
only utilize Weyl and Ising fermions. Any rank reduction
is due to Ising fermions. These examples sample symmetric
orbifold models.

$\bullet$ NAHE: each generation comes from a distinct sector, with
different left and right-moving structure.

$\bullet$ Models 1-3: two generations come from distinct sectors.
The third comes from a sector which differs
from one of these sectors
only in
its left-moving spin structure.

\noindent{\it Real fermionic realizations:} these solutions
utilize Weyl and chiral Ising fermions. In Models 4,5
all of the chiral Ising fermions are left-moving. These
examples are likely to have an asymmetric orbifold interpretation.
In Models 6,7 four of the chiral Ising fermions are right-movers.
These examples belong to the general class of asymmetric $(0,2)$
Gepner constructions.

$\bullet$ Model 4: two generations come from distinct sectors, but the
third generation comes from a sector which is effectively
the sum of the first two sectors:
adding sector 1, which contains the gravitino, simply
takes the scalar component of a supermultiplet into the fermionic
component, while taking sector 5 to 3*5 takes states into
their CPT conjugates within the same multiplet.

$\bullet$ Model 5: two generations come from the same sector; the
third lives in a distinct sector. In fact,
the quark doublets of the two generations in the same sector
differ only by a single $U(1)$ charge of $G_{flavor}$.

$\bullet$ Model 6: similar to Models 1-3.

$\bullet$ Model 7: similar to Model 5.

These results are encouraging for semi-realistic model building.
As mentioned in the introduction, the real fermion construction
samples the full range of exactly solvable $(0,2)$ constructions.
We have found examples with precisely three chiral generations
in every such class, and with generic gauge group and matter
content.

\subsec{Hypercharge and $k_1$}

As mentioned in section 2 our construction explores
the {\it generic} class of embeddings of the standard model
gauge group allowed in string theory, as opposed to
embeddings in grand unified groups like $SU(5)$ or $E_6$.
In the fermionic construction, this is achieved by
exploring overlapping embeddings of the standard model
gauge group with embeddings of $G_{flavor}$ and $G_{hidden}$
in the fermionic charges. A consequence is to increase
the number of distinct hypercharge embeddings
which are possible for each conformal field theory solution
we construct. In fact for some solutions
there are one parameter continously varying
embeddings of the hypercharge. As a result,
$k_1$ can be {\it continuously variable} within the same
conformal field theory solution.

We emphasize that this is merely a statement about the
flexibility in hypercharge embedding and particle
identification within these solutions, and should
not be misinterpreted as the continuous variation
of $k_1$ {\it within} a string model.
In fact, a general theorem in perturbative string theory
tells us that there is no continuously varying modulus
that can adjust the value of $k_1$ in an
$N$$=$$1$ heterotic string model with chiral matter
\refs{\bdfm}. Our results are consistent with this theorem.
A given hypercharge embedding fixes both $k_1$ and much of the
particle identification in the conformal
field theory solution, thus {\it defining}
a string model up to the vacuum shift necessary for
removing the anomalous $U(1)$.

We determine an acceptable
choice for $k_1$ in two steps.
To find a consistent
hypercharge embedding,
we solve for a nonanomalous $U(1)$ for which
we can identify a full 3 generations of quarks and leptons
with conventional values of hypercharge in the massless
spectrum. Since this definition says nothing about the Higgs,
one must then check in each case whether or not there
appear a pair of candidate electroweak Higgs doublets with
conventional hypercharge. At this stage, for e.g.
in Model 6, it is still possible that a continuous range of $k_1$
values is allowed.  We now determine the hypercharge
embedding by {\it requiring} that it is defined so as to minimize
the number of fractionally charged exotics, also avoiding
any hypercharge mismatch in what would otherwise be
pairs of vectorlike exotics.

Let us see how this works out in particular examples.
Consider Model 4, which we see from Table 2.1 has 5
nonanomalous $U(1)$ generators: $Q_1$-$Q_5$. As far as
obtaining three generations of standard model quarks and leptons,
there are 5 possible definitions of hypercharge:
$$\eqalign{
Y_1 &= {1\over 3}\left( -{3\over 20}Q_1 + {27\over 320}Q_2
+ {3\over 40}Q_3 + {1\over 24}Q_4 - {1\over 24}Q_5
+ {5\over 192}Q_6 \right)\cr
Y_2 &= {1\over 3}\left( -{3\over 5}Q_1 - {9\over 40}Q_2
+ {3\over 10}Q_3 + {1\over 6}Q_4 - {1\over 6}Q_5
- {1\over 12}Q_6 \right)\cr
Y_3 &= {1\over 3}\left( -{9\over 10}Q_1 - {21\over 160}Q_2
- {3\over 20}Q_3 - {1\over 12}Q_4 - {7\over 60}Q_5
+ {1\over 96}Q_6 \right)\cr
Y_4 &= {1\over 3}\left( {3\over 10}Q_1 + {69\over 160}Q_2
- {3\over 5}Q_3 - {1\over 3}Q_4 + {11\over 60}Q_5
+ {7\over 96}Q_6 \right)\cr
Y_5 &= {1\over 3}\left( {9\over 10}Q_1 + {39\over 160}Q_2
+ {3\over 10}Q_3 + {1\over 6}Q_4 + {1\over 12}Q_5
- {11\over 96}Q_6 \right)\cr
}$$

All 5 choices also provide at least one pair of candidate
electroweak Higgs doublets, so this criterion does not
distinguish between them. Two of these choices, $Y_1$ and $Y_3$, give a
reduced spectrum of fractionally charged states. All of these
choices except $Y_1$ have the unpleasant feature that the two exotic
color triplets are not truly vectorlike, i.e. their hypercharges
are not equal and opposite to those of the two extra color antitriplets.
The $k_1$ values associated with these 5 choices are given
respectively by
\eqn\konesix{
k_1 = {35\over 24},\;{46\over 3},\;{29\over 6},\;
{185\over 6},\;{125\over 6}
}
All of these values are quite large except for the first
one, associated with $Y_1$.
Incidently, $k_1$$=$$35/24$ is $12.5\%$
less than $k_1$$=$$5/3$, a value close to optimal
for the scenario of improving gauge coupling unification by
adjusting $k_1$.

So in the case of Model 4 we are quickly led to a unique
choice of hypercharge, once we impose some phenomenological
criteria.

Another interesting example is Model 6. We see from
Table 2.3 that it has 10 nonanomalous $U(1)$ generators,
however the first three of these clearly belong to
$G_{hidden}$. Let us call the remaining seven $Q_1$-$Q_7$.
There is considerable choice in the hypercharge embedding
for this conformal field theory solution. Among the allowed
possibilities is the following
one-parameter set:

$$\eqalign{
Y_3 &= {1\over 3}\big( -{15\over 16}(1+24q)Q_1 - {3\over 4}Q_2
+ {3\over 20}Q_3 -{2\over 5}(1+10q)Q_4 \cr
&+ {5\over 16}(1+24q)Q_5 + {1\over 40}(7+160q)Q_6 + qQ_7 \big)\cr
}$$
where $q$ is an arbitrary real parameter.

The corresponding one-parameter set of
possible $k_1$ values are given by:
\eqn\kval{
k_1 = {103\over 12} + 260q +2800q^2
}
There are 9 weak doublet states in this conformal field
theory solution from
which we must identify candidates for the
up and down type
electroweak Higgs. The hypercharges of some
(though not all) of these doublets depend on $q$.
There are also three pairs of vectorlike color triplet
exotics, some of whose hypercharges also depend on $q$.
The hypercharges of the 9 weak doublets are given by:
$3$, $3$, $-$$3$, $9$, $9$, $(6$$+$$120q)$, $-$$(6$$+$$120q)$,
$(15$$+$$240q)$, $-$$(15$$+$$240q)$. The hypercharges of the
three color triplet exotics are $4$, $-$$(5$$+$$120q)$,
and $(10$$+$$240q)$, while those of the color antitriplet
exotics are $-4$, $-4$, and $(5$$+$$120q)$.
We are led to a unique choice,
$q$$=$$-1/40$, to avoid a hypercharge mismatch for the color
triplet exotics. The corresponding value of $k_1$ is $23/6$.
Table 2.3 gives the spectrum corresponding to this hypercharge
embedding.

\subsec{Model 5}

Rather than go into details for all of our conformal
field theory solutions we will
be content in the remainder to focus on one model,
obtained from the solution described in Table 2.2. From
Table 2.2 we see that there are 6 nonanomalous
$U(1)$'s: $Q_1$-$Q_6$.
There are 4 possible hypercharge definitions for this
conformal field theory solution, but by the same procedure
as above we are quickly
led to a unique choice:
\eqn\yseven{
Y = {1\over 48}\left( -8Q_2 - 3Q_3 - 8Q_4 - Q_5 + Q_6 \right)
}
The corresponding value of $k_1$ is $11/6$, which we
henceforth refer to as Model 5. This is interesting,
as it is only slightly larger than the $SU(5)$ value $5/3$.

It is interesting that the
perturbative heterotic string vacuum corresponding
to Model 5 can be obtained from two distinct fermionic realizations.
The basis vectors corresponding to these two
different embeddings are given in Tables A.5 and A.8.
As shown in the appendix,
the gauge embeddings of $SU(3)_c$ and $SU(2)_L$
in free fermionic charges are different in these two realizations.
Nevertheless we have verified that the massless spectra are
identical, and have checked that the superpotentials agree
at least through quintic order. This demonstrates that
the free fermionic realization of the gauge embedding
is {\it not} an invariant property of the cft solution.

The second version of Model 5, Table A.8, has the property
that by simply removing the final basis vector we obtain
a model in which $SU(3)_c$$\times$$SU(2)_L$ is promoted
to $SU(5)$. Thus we may ask the question: what happened
to the conventional $k_1$$=$$5/3$ hypercharge
$U(1)_Y$$\subset$$SU(5)$ when we broke $SU(5)$ to
$SU(3)_c$$\times$$SU(2)_L$? The answer is that the
fermionic charge vector which would correspond to this
$U(1)_Y$ is not orthogonal to the roots of the level two
$SU(4)$ hidden sector group. Thus we do not quite have
an $SU(5)$ based embedding, since the true hypercharge must involve
a mixture with the other abelian generators.
Nevertheless the actual value of $k_1$, $11/6$, is quite close
to the $SU(5)$ value.

A complete listing of the nonvanishing terms in
the cubic, quartic, and quintic effective superpotential of Model 5
is given in Tables 4.1-4.3. As described in section 2.5,
the numerical values of the couplings can, with some effort,
be computed; they are generically order one.

Let us examine how the possible patterns of quark, lepton,
and Higgs masses are related to the vacuum shift associated
with the anomalous $U(1)$. We immediately observe from
the cubic superpotential a term $Q_3\,u^c_3\,\bar{h}_3$, which
can be interpreted as giving mass to the top quark, provided
that $\bar{h}$ remains light and can serve as the up type
electroweak Higgs. We then examine, at the cubic level,
the full Higgs mass matrix, including mixings with $L_2$ and
$L_3$:
\eqn\higgs{
M = \bordermatrix{&h_1&h_2&h_3&h_4&L_2&L_3\cr
\bar{h}_1&0&0&\chi_9&0&0&0\cr
\bar{h}_2&\chi_2&0&0&0&\chi_{11}&0\cr
\bar{h}_3&0&\chi_4&0&0&\chi_6&\chi_5\cr
\bar{h}_4&0&\chi_{12}&0&0&0&\chi_{13}\cr
}
}
By diagonalizing $MM^{\dagger}$ and $M^{\dagger}M$ we
can tell which combination of fields remain light
when various entries in $M$ become large after the vacuum shift.
As one would expect, $\bar{h}_3$ does not remain light
unless neither $\chi_4$, $\chi_5$, nor $\chi_6$
gets a superheavy vev. In this case both $L_2$ and $L_3$
remain light, but mix with $h_2$.

A reasonable requirement we could make on the vevs
at this level is that two pairs of up and down type Higgs
should remain light after the vacuum shift.
Note that the vectorlike
color triplet exotic pair remains light after the vacuum
shift. An extra light Higgs pair would fill out
an approximate $5$,$\bar{5}$ of $SU(5)$, minimizing
the effect of the color exotics on the RG running of the
gauge couplings \lang .
One solution to this requirement is that
$\chi_2$ and $\chi_{11}$ should
not get superheavy vevs.

This leaves $\chi_9$, $\chi_{12}$, and $\chi_{13}$ to
get vevs in the vacuum shift. We must then ask whether there
is a set of vevs which includes these fields, cancels the
Fayet-Iliopoulos term \dflat, and satisfies all of
the F and D flatness conditions \fflat. A simple
solution does exist: the vacuum shift involves appropriate
vevs for the fields
\eqn\vacset{
\{ \chi_9,\;\chi_{12},\;\chi_{13},\;\chi_{14},\;T_1,\;T_2,\;T_3 \}
}
By examining Table 2.2 one sees that this set satisfies the
D flatness conditions; we have checked that it satisfies the
F flatness conditions at least through 8th order in the
superpotential.

After this vacuum shift, $\bar{h}_3$, $\bar{h}_2$, $h_1$, and
$h_4$ remain massless. In addition, $L_2$ remains massless,
as does the combination
$<$$\chi_{12}$$>$$L_3 - $$<$$\chi_{13}$$>$$h_2$.
This shift also breaks the level two $SU(2)$ of $G_{hidden}$.
If we proceed to the quartic level in this scenario,
we notice the terms
\eqn\btau{
Q_3\,d^c_3\,h_1\,\chi_{12} + L_3\,e^c_3\,h_1\,\chi_{12}
}
which give mass to the bottom quark and tau lepton.
These masses are suppressed by $<$$\chi_{12}$$>$$/$$M_{str}$
relative to the top mass.
If $<$$\chi_{13}$$>$$/$$<$$\chi_{12}$$>$ is not too large, we
also reproduce approximate $b$-$\tau$ Yukawa unification
as in $SU(5)$.

In this model there are no dangerous
B violating terms through quintic order, provided that
$\phi_{12}$ does not get a superheavy vev. This statement
is somewhat dependent on particle identification, but
in any event it does not seem difficult
to avoid rapid proton decay.
To have a truly viable model, we would also need
large masses for the other exotics, and higher
order mass terms for (at least most of)
the 1st and 2nd generation quarks
and leptons (as well as mixings).

\newsec{Conclusions}

Arriving at precisely three generations of massless chiral fermions
had proven to be a notoriously unpredictable step in superstring model
building. Prior to our work, the only three generation models
for which both the
massless spectrum and the superpotential have been computed are
symmetric orbifolds with Wilson lines, which includes the free fermionic
examples. The models in this paper go beyond that class.
They are the first known examples of three generation models based on
genuinely heterotic modular invariants, obtained by tensoring together
holomorphic cfts which are not free fields. We expect this feature will be
generic to other exactly solvable $(0,2)$ constructions, suggesting
that there exist many new starting points for obtaining three generations.

The phenomenology of these models needs to be worked out in
detail and compared with that of previously known models.
Such an analysis will suggest new strategies for model
building. It would be helpful to have a better understanding
of string threshold effects in order to to make progress
on the problem of gauge coupling unification.
It is certainly possible to investigate the problem of
exotics in string models embedding the MSSM.
Since the number of exotics is variable, perhaps
it can be reduced to zero, leaving just the MSSM
matter content in the visible sector. In this event,
it would be important to understand whether string threshold
effects can be large giving an alternative explanation for
the mismatch between the unification scales. The question
of moduli dependence of such thresholds has been
investigated in orbifold constructions with differing
conclusions from those models investigated in the
free fermionic construction \refs{\niland,\bachas}. It
would also be nice to reduce the size of $G_{flavor}$, and
perhaps make contact with various texture schemes \refs{\binetruy}.

A common feature of all known superstring embeddings of the
MSSM is the presence of extra low-energy matter. It
is intriguing that every semi-realistic example to date also has
a tree-level anomalous $U(1)$.  However, it is not known whether
these are {\it essential} features of a heterotic string vacuum
embedding the MSSM. In order to address this question convincingly,
one must explore a {\it large sample} of superstring embeddings
of the MSSM. The examples in this paper are only a beginning in
this direction, but they sample a wide range of exactly
solvable $(0,2)$ constructions.

We should mention that we have found exceptions without these
features in admittedly unrealistic three generation models. We have
an example of a $SO(10)$ level one model with precisely three chiral
$\bf 16$'s
and {\it no } additional vectorlike matter transforming
under the $SO(10)$.
Interestingly, a slight change in the fermionic construction of this
solution converts it into a three generation model with $SU(5)$ realized
at level two plus an exotic chiral $\bf 15$, hidden sector group $F_4$,
flavor group $(U(1))^6$ but {\it no} anomalous $U(1)$.

Finally, we note that there are questions of phenomenological
interest which would be most easily explored in pedagogical
models with, for example, a {\it single} chiral generation of
quarks and leptons. We can construct many such examples.
Given the superpotential with both singlet couplings and
nonrenormalizable terms included, one could investigate the
absence of the $\mu$ term in the superpotential or look for
specific couplings necessary for generating an intermediate scale.

\centerline{\bf Acknowledgements}

We would like to thank the participants of the Planck'95
workshop at the ITP for many discussions,
especially P. Langacker, P. Binetruy, M. Cvetic,
V. Kaplunovsky, J. Louis, H.-P. Nilles,
J. Polchinski, R. Schimmrigk, and M. Srednicki. We also
thank K. Dienes and A. Faraggi for discussions of their
work. This work was supported in part by National Science
Foundation grants PHY 91-16964 and PHY 94-07194.

\listrefs
\appendix{A}{Details of the fermionic embeddings}

In this appendix we briefly describe the specifications
of the seven solutions discussed in this paper. The real fermionic
construction is described in \refs{\klst ,\cchl}. Tables A.1
through A.7 list the basis vectors and $k_{ij}$ matrices that
define each solution. Table A.8 gives the equivalent version
of Model 5.
Our definition of $k_{ij}$ is as in \klst .
Each basis vector specifies 20 right-moving and 44 left-moving
fermion boundary conditions in Majorana-Weyl notation.
A double vertical line separates the right-movers from the
left-movers; a single vertical line is used to separate the
left-moving Weyl fermion boundary conditions from those of
the left-moving Ising or chiral Ising fermions. A ``0'' denotes
Neveu-Schwarz boundary conditions, while a ``1'' denotes Ramond.
The detailed map between our notation for boundary conditions
and that of \klst\ is given by:

$$\eqalign{
0 &: 0;\qquad\quad 1 : -1/2;\qquad + : +1/4;\qquad - : -1/4;\cr
e &: 1/8;\qquad x : -1/8;\qquad t : 3/8;\cr
}$$

The simple roots of $SU(3)_c$$\times$$SU(2)_L$ are embedded
in the first 8 fermionic charges of the left-movers. This embedding
is the same in all seven solutions:

$$\eqalign{
&SU(3)_c :\cr
&\b(\ha ,\ha ,\mha ,\mha ,\mone ,\ze ,\ze ,\ze ,\ze ,\ze ,\ze\e)\cr
&\b(\ze ,\ze ,\ze ,\one ,\ha ,\ha ,\mha ,\ha ,\ze ,\ze ,\ze\e)\cr
&SU(2)_L :\cr
&\b(\ha ,\ha ,\mha ,\mha ,\one ,\ze ,\ze ,\ze ,\ze ,\ze ,\ze\e)\cr
}$$

We note that in the equivalent fermionic realization
of Model 5 defined by Table A.8, the embedding of
$SU(3)_c$$\times$$SU(2)_L$ is quite different:

$$\eqalign{
&SU(3)_c :\cr
&\b(\qt ,\mqt ,\mqt ,\mtq ,\ha ,\ze ,\mha ,\ze ,\ha ,\ze ,
\ha ,\mha ,\ze ,\ze\e)\cr
&\b(\ze ,\ze ,\one ,\one ,\ze ,\ze ,\ze ,\ze ,\ze ,\ze ,
\ze , \ze ,\ze ,\ze\e)\cr
&SU(2)_L :\cr
&\b(\ha ,\mha ,\ha ,\mha ,\ze ,\ze ,\ze ,\ze ,\ze ,\ze ,
\ze ,\one ,\ze ,\ze\e)\cr
}$$

\medskip
The embedding of the simple roots of
the nonabelian factors of $G_{hidden}$ into the fermionic charges of the
left-movers is model dependent. We list these embeddings below
for all seven solutions:

\goodbreak\medskip\medskip
\noindent Model 1:

$$\myeqalign{
&SU(2)_2 :\cr
&\b(\ze ,\ze ,\ze ,\ze ,\ze ,\ze ,\ze ,\ze ,\ze ,\ze ,
\ze ,\one ,\ze ,\ze ,\ze ,\ze ,\ze ,\ze ,\ze ,\ze\e)\cr
&SU(2) :\cr
&\b(\ze ,\ze ,\ze ,\ze ,\ze ,\ze ,\ha ,\ha ,\ze ,\mha ,\ha ,
\ze ,\ze ,\ze ,\ze ,\ze ,\ze ,\ze ,\mone ,\ze\e)\cr
&SU(2) :\cr
&\b(\ze ,\ze ,\ze ,\ze ,\ze ,\ze ,\ha ,\ha ,\ze ,\mha ,
\ha ,\ze ,\ze ,\ze ,\ze ,\ze ,\ze ,\ze ,\one ,\ze\e)\cr
}$$

\goodbreak\medskip
\noindent Model 2:

$$\myeqalign{
&SO(7) :\cr
&\b(\ze ,\ze ,\ze ,\ze ,\ze ,\ze ,\ha ,\ha ,\ze ,\mha ,
\ha ,\mone ,\ze ,\ze ,\ze ,\ze ,\ze ,\ze ,\ze ,\ze\e)\cr
&\b(\ze ,\ze ,\ze ,\ze ,\ze ,\ze ,\ze ,\ze ,\ze ,\ze ,
\ze ,\one ,\ze ,\ze ,\ze ,\ze ,\ze ,\ze ,\mone ,\ze\e)\cr
&\b(\ze ,\ze ,\ze ,\ze ,\ze ,\ze ,\ze ,\ze ,\ze ,\ze ,
\ze ,\ze ,\ze ,\ze ,\ze ,\ze ,\ze ,\ze ,\one ,\ze\e)\cr
&SU(2)_2 :\cr
&\b(\ze ,\ze ,\ze ,\ze ,\ze ,\ha ,\ha ,\ze ,\ha ,\ha ,\ze ,
\ze ,\ze ,\ze ,\ze ,\ze ,\ze ,\ze ,\ze ,\ze\e)\cr
&SU(2)_2 :\cr
&\b(\ze ,\ze ,\ze ,\ze ,\ze ,\ze ,\ze ,\ze ,\ze ,\ze ,
\ze ,\ze ,\ze ,\ze ,\ze ,\ze ,\ze ,\ze ,\ze ,\one\e)\cr
&SU(2) :\cr
&\b(\ze ,\ze ,\ze ,\ze ,\ze ,\ze ,\ze ,\ze ,\ze ,\ze ,
\ze ,\ze ,\ze ,\ze ,\ze ,\ze ,\one ,\mone ,\ze ,\ze\e)\cr
&SU(2) :\cr
&\b(\ze ,\ze ,\ze ,\ze ,\ze ,\ze ,\ze ,\ze ,\ze ,\ze ,
\ze ,\ze ,\ze ,\ze ,\one ,\one ,\ze ,\ze ,\ze ,\ze\e)\cr
&SU(2) :\cr
&\b(\ze ,\ze ,\ze ,\ze ,\ze ,\ze ,\ze ,\ze ,\ze ,\ze ,
\ze ,\ze ,\ze ,\ze ,\one ,\one ,\ze ,\ze ,\ze ,\ze\e)\cr
&SU(2) :\cr
&\b(\ze ,\ze ,\ze ,\ze ,\ze ,\ze ,\ze ,\ze ,\ze ,\ze ,
\ze ,\ze ,\ze ,\ze ,\ze ,\ze ,\one ,\one ,\ze ,\ze\e)\cr
}$$

\goodbreak\medskip
\noindent Model 3:

$$\myeqalign{
&SO(5) :\cr
&\b(\ze ,\ze ,\ze ,\ze ,\ze ,\ze ,\ha ,\ha ,\ze ,\mha ,
\ha ,\mone ,\ze ,\ze ,\ze ,\ze ,\ze ,\ze ,\ze ,\ze\e)\cr
&\b(\ze ,\ze ,\ze ,\ze ,\ze ,\ze ,\ze ,\ze ,\ze ,\ze ,
\ze ,\one ,\ze ,\ze ,\ze ,\ze ,\ze ,\ze ,\ze ,\ze\e)\cr
&SO(5) :\cr
&\b(\ze ,\ze ,\ze ,\ze ,\ze ,\ha ,\ha ,\ze ,\ha ,\ha ,\ze ,
\ze ,\ze ,\ze ,\ze ,\ze ,\mone ,\ze ,\ze ,\ze\e)\cr
&\b(\ze ,\ze ,\ze ,\ze ,\ze ,\ze ,\ze ,\ze ,\ze ,\ze ,\ze ,
\ze ,\ze ,\ze ,\ze ,\ze ,\one ,\ze ,\ze ,\ze\e)\cr
&SU(2)_2 :\cr
&\b(\ze ,\ze ,\ze ,\ze ,\ze ,\ze ,\ze ,\ze ,\ze ,\ze ,
\ze ,\ze ,\ze ,\ze ,\one ,\ze ,\ze ,\ze ,\ze ,\ze\e)\cr
}$$

\goodbreak\medskip
\noindent Model 4:

$$\myeqalign{
&SU(4)_2 :\cr
&\b(\ze ,\ze ,\ze ,\ze ,\ze ,\ha ,\ze ,\mha ,\ha ,\ze ,
\mha ,\ze ,\ze ,\ze\e)\cr
&\b(\qt ,\qt ,\qt ,\qt ,\ze ,\mha ,\ze ,\ze ,\mha ,\ha ,
\ze ,\ze ,\ze ,\ze\e)\cr
&\b(\ze ,\ze ,\ze ,\ze ,\ze ,\ha ,\ha ,\ze ,\ze ,\mha ,
\ha ,\ze ,\ze ,\ze\e)\cr
&SU(2)_2 :\cr
&\b(\ze ,\ze ,\ze ,\ze ,\ze ,\ze ,\ze ,\ze ,\ze ,\ze ,
\ze ,\ze ,\ze ,\one\e)\cr
}$$

\goodbreak\medskip
\noindent Model 5:

$$\myeqalign{
&SU(4)_2 :\cr
&\b(\ze ,\ze ,\ze ,\ze ,\ze ,\ha ,\ze ,\mha ,\ha ,\ze ,
\mha ,\ze ,\ze ,\ze\e)\cr
&\b(\qt ,\qt ,\qt ,\qt ,\ze ,\mha ,\ze ,\ze ,\mha ,\ha ,
\ze ,\ze ,\ze ,\ze\e)\cr
&\b(\ze ,\ze ,\ze ,\ze ,\ze ,\ha ,\ha ,\ze ,\ze ,\mha ,
\ha ,\ze ,\ze ,\ze\e)\cr
&SU(2)_2 :\cr
&\b(\ze ,\ze ,\ze ,\ze ,\ze ,\ze ,\ze ,\ze ,\ze ,\ze ,
\ze ,\one ,\ze ,\ze\e)\cr
}$$

\goodbreak\medskip
\noindent Model 6:

$$\myeqalign{
&SU(2)_2 :\cr
&\b(\ze ,\ze ,\ze ,\ze ,\ze ,\ha ,\ha ,\ze ,\ha ,\ha ,
\ze ,\ze ,\ze ,\ze\e)\cr
&SU(2)_2 :\cr
&\b(\ze ,\ze ,\ze ,\ze ,\ze ,\ze ,\ha ,\ha ,\ze ,\mha ,
\ha ,\ze ,\ze ,\ze\e)\cr
}$$

\goodbreak\medskip
\noindent Model 7:

$$\myeqalign{
&SO(7) :\cr
&\b(\ze ,\ze ,\ze ,\ze ,\ze ,\ha ,\ha ,\ze ,\ha ,\ha ,
\ze ,\ze ,\ze ,\ze ,\mone ,\ze\e)\cr
&\b(\ze ,\ze ,\ze ,\ze ,\ze ,\ze ,\ze ,\ze ,\ze ,\ze ,
\ze ,\ze ,\ze ,\ze ,\one ,\mone\e)\cr
&\b(\ze ,\ze ,\ze ,\ze ,\ze ,\ze ,\ze ,\ze ,\ze ,\ze ,
\ze ,\ze ,\ze ,\ze ,\ze ,\one\e)\cr
&SU(2)_2 :\cr
&\b(\ze ,\ze ,\ze ,\ze ,\ze ,\ze ,\ze ,\ze ,\ze ,\ze ,\ze ,
\one ,\ze ,\ze ,\ze ,\ze\e)\cr
}$$

Having defined the embeddings of the nonabelian factors into
the left-moving fermionic charges, the $U(1)$ embeddings are
defined to span the remaining
$n$ dimensional orthogonal subspace, where $n$ varies from 7
to 14 in our solutions.
There is
of course a great deal of freedom in the choice of a basis
for these $U(1)$ embeddings. Our computer program chooses
a basis by first identifying the anomalous U(1); this is then
designated as $U_{n-1}$. A basis is then chosen for the
remaining $n$$-$$1$ nonanomalous $U(1)$'s, $U_0$--$U_{n-2}$,
such that the complete
$U(1)$ basis is orthogonal. The basis is chosen such that all
of the elements are integers, and the program attempts to
minimize the norms, $\vert U_i\vert^2$,
of all the basis vectors.
The $U(1)$ charges listed in Tables 2.1 to 2.3 are obtained
as follows: if $\vec f$ is the vector of
left-moving fermionic charges
of a state, then the $(i$$+$$1)$st $U(1)$ charge listed in the
table would be $4\vec f\cdot\vec U_i$. The additional factor of
4 ensures that all of the charge entries will be integers.

Below we give the $U(1)$ embedding basis,
$U_0$--$U_{n-1}$, for each solution. Keep in mind that
the last vector listed in each case defines the anomalous $U(1)$.
The integer just to the right of each basis vector is its norm;
we must keep track of these in order to, for example, compute $k_1$.

{\obeylines

Model 1:

 0: (  0  0  0  0  0  0  0  0  0  0  0  0  0  0  1  0  0  0  0  0) 1
 1: (  0  0  0  0  0  0  0  0  0  0  0  0  0  0  0  1  0  0  0  0) 1
 2: (  0  0  0  0  0  0  0  0  0  0  0  0  0  0  0  0  1  0  0  0) 1
 3: (  0  0  0  0  0  0  0  0  0  0  0  0  0  0  0  0  0  1  0  0) 1
 4: (  0  0  0  0  0  0  0  0  0  0  0  0  0  0  0  0  0  0  0  1) 1
 5: (  0  0  0  0  0 -4 -2  2  0  0  0  0  0 -3  0  0  0  0  0  0) 33
 6: (  0  0  0  0  0  8  4 -4  0  0  0  0 11-16  0  0  0  0  0  0) 473
 7: (  5 -5  0  0  0  0  0  0  6  0  0  0  0  0  0  0  0  0  0  0) 86
 8: (  3 -3  0  0  0 -2 -1  1 -5  0  0  0  8  4  0  0  0  0  0  0) 129
 9: ( 31 55 86  0  0 -4 -2  2 20  0  0  0 16  8  0  0  0  0  0  0) 12126
10: (-16  8 -8  0  0 -4 45 49 20 47-47  0 16  8  0  0  0  0  0  0) 9964
11: ( 28-14 14  0  0  7 14  7-35 90 69  0-28-14  0  0  0  0  0  0) 16536
12: (  1 19-19 39  0-29 20-29 15  6 15  0 12  6  0  0  0  0  0  0) 4992
13: (  5 -1  1  3  0 -1  4 -1 -5 -2 -5  0 -4 -2  0  0  0  0  0  0) 128

Model 2:

 0: (  0  0  0  0  0  0  0  0  0  0  0  0  1 -1  0  0  0  0  0  0) 2
 1: (  0  0  0  0  0 -1  0  1  1  0 -1  0  0  0  0  0  0  0  0  0) 4
 2: (  0 -1 -1  0  0  0  0  0  0  0  0  0  0  0  0  0  0  0  0  0) 2
 3: (  1  0  0  1  0 -1  0 -1  1  0  1  0  0  0  0  0  0  0  0  0) 6
 4: ( -6  4 -4  2  0  0  1 -3  2 -3 -1  0 -6 -6  0  0  0  0  0  0) 168
 5: ( -1  1 -1  1  0  0  1 -1  0 -1 -1  0  2  2  0  0  0  0  0  0) 16
 6: ( -1  1 -1  1  0 -2  1  1 -2  3  1  0  0  0  0  0  0  0  0  0) 24
 7: (  5 -1  1  3  0  0  5 -1 -4 -1 -5  0 -2 -2  0  0  0  0  0  0) 112

Model 3:

 0: (  0  0  0  0  0  0  0  0  0  0  0  0  0  0  0  1  0  0  0  0) 1
 1: (  0  0  0  0  0  0  0  0  0  0  0  0  0  0  0  0  0  1  0  0) 1
 2: (  0  0  0  0  0  0  0  0  0  0  0  0  0  0  0  0  0  0  1  0) 1
 3: (  0  0  0  0  0  0  0  0  0  0  0  0  0  0  0  0  0  0  0  1) 1
 4: (  0  0  0  0  0  0  0  0 -1  1  1  0  0 -1  0  0  0  0  0  0) 4
 5: (  0 -1 -1  0  0  0  0  0  0  0  0  0  0  0  0  0  0  0  0  0) 2
 6: ( -2  1 -1  0  0  0  0  0  0  0  0  0 -2  0  0  0  0  0  0  0) 10
 7: (  0  0  0  0  0 -1  1  2 -3  3  0  0  0  6  0  0  0  0  0  0) 60
 8: (  0  0  0  0  0  2  0 -2 -1 -1  1  0  0  1  0  0  0  0  0  0) 12
 9: ( -4  2 -2  0  0  2  3  1 -4 -1 -5  0  6 -2  0  0  0  0  0  0) 120
10: (  1  7 -7 15  0-12  7-11  4  1  5  0  6  2  0  0  0  0  0  0) 720
11: (  5 -1  1  3  0  0  5 -1 -4 -1 -5  0 -6 -2  0  0  0  0  0  0) 144

\goodbreak
Model 4:

 0: (  0  0  0  0  0  0  0  0  0  0  0  1  2  0) 5
 1: (  6 -6  0  0  0  0  1  1 -1 -1 -2  8 -4  0) 160
 2: (  2 -2  0  0  0  0 -3 -3 -2 -2  1 -2  1  0) 40
 3: (  0  0 -2  2  0 -2 -1 -3  4  2  5  2 -1  0) 72
 4: (  3 -3  5 -5  0  5 -2  3  2  7  4 -2  1  0) 180
 5: (  6 18 28 -4  0 -8-13  3 13-19  2  8 -4  0) 2016
 6: (  4 -2  0  2  0 -3  3  2  4 -1 -1 -4  2  0) 84

Model 5:

 0: (  0  0  0  0  0  0  0  0  0  0  0  0  0  1) 1
 1: (  1 -1  0  0  0  0  0  0  0  0  0  0  1  0) 3
 2: ( -2  2  0  0  0  0 -1 -1 -3 -3 -2  0  4  0) 48
 3: (  0  0  0  0  0  0  1  1  0  0 -1  0  0  0) 3
 4: (  1 -1  3 -3  0  3 -2  1 -3  0 -1  0 -2  0) 48
 5: (  5 -1  1  3  0 -5 -1 -2  0 -9 -3  0 -6  0) 192
 6: (  1  3  5 -1  0 -1 -1  2  4 -1  1  0  2  0) 64

Model 6:

 0: (  0  0  0  0  0  0  0  0  0  0  0  1  0  0  0  0) 1
 1: (  0  0  0  0  0  0  0  0  0  0  0  0  0  0  1  0) 1
 2: (  0  0  0  0  0  0  0  0  0  0  0  0  0  0  0  1) 1
 3: (  0  0  0  0  0  0  0  0 -1  1  1  0  0 -1  0  0) 4
 4: (  0 -1 -1  0  0  0  0  0  0  0  0  0  0  0  0  0) 2
 5: ( -2  1 -1  0  0  0  0  0  0  0  0  0 -2  0  0  0) 10
 6: (  0  0  0  0  0 -1  1  2 -3  3  0  0  0  6  0  0) 60
 7: (  0  0  0  0  0  2  0 -2 -1 -1  1  0  0  1  0  0) 12
 8: ( -4  2 -2  0  0  2  3  1 -4 -1 -5  0  6 -2  0  0) 120
 9: (  1  7 -7 15  0-12  7-11  4  1  5  0  6  2  0  0) 720
10: (  5 -1  1  3  0  0  5 -1 -4 -1 -5  0 -6 -2  0  0) 144

\goodbreak
Model 7:

 0: (  0  0  0  0  0  0  0  0  0  0  0  0  1  0  0  0) 1
 1: (  0  0  0  0  0  0  0  0  0  0  0  0  0  1  0  0) 1
 2: (  0  0  0  0  0  0  0  0  1 -1  0  0  0  0  0  0) 2
 3: (  0  1  1  0  0  0  0  0  0  0  0  0  0  0  0  0) 2
 4: (  0  0  0  0  0 -1 -1  0  1  1  2  0  0  0  0  0) 8
 5: (  1  1 -1  3  0 -1  3 -2 -1 -1  2  0  0  0  0  0) 32
 6: ( -1  1 -1  1  0 -1 -1 -2  1  1 -2  0  0  0  0  0) 16
 7: (  2 -1  1  0  0  1 -1 -2  0  0  0  0  0  0  0  0) 12
 8: (  3 -1  1  1  0 -3  1  2  1  1 -2  0  0  0  0  0) 32

}

\listrefs
%\end
\vfil\eject
\centerline{TABLE CAPTIONS}
\vskip 1in
\leftline{TABLE 1:}
\noindent Summary of the 7 three generation solutions.
The first column lists the amount
of rank reduction $c_m$, where $22$$-$$c_m$ is
the rank of the full gauge group.
The second column lists
the number, $n$, of nonanomalous $U(1)$'s in
$G_{flavor}$. The third column lists
the number of vectorlike pairs of
color triplet exotics, while the fourth column lists
the number of extra weak doublets which are also
singlets under the nonabelian part of $G_{hidden}$. The
fifth column gives
$G_{hidden}$, and the last column
indicates whether or not there is chiral matter
transforming under $G_{hidden}$.

\smallskip
\leftline{TABLES 2.1-2.3:}
\noindent The complete list of massless chiral superfields for
the conformal field theory solutions which give Models 4,5, and 6,
respectively. This is the light spectrum {\it before} the
vacuum shift required by the presence of an anomalous $U(1)$.
The row of integers listed for each state are its charges under
the full set of nonanomalous $U(1)$'s and the anomalous $U(1)$.
The charge under the anomalous $U(1)$ is the last one listed
for each state. The normalization and choice of orthogonal basis
for the $U(1)$'s is discussed in the appendix. The hypercharges
of the states are indicated separately; the embedding of the
hypercharge into the nonanomalous $U(1)$'s is given in sections
4.2 and 4.3 of the text.
\smallskip
\leftline{TABLE 3:}
\noindent For each model, we indicate the three sectors
of the fermionic realization
containing the
positive helicity fermionic component highest weight
states of the three generations of chiral superfields.
The sectors are listed as linear
combinations of the basis vectors for each fermionic cft solution;
the basis vectors
are given in Tables A.1 to A.7.
\smallskip
\leftline{TABLES 4.1-4.3:}
\noindent The complete cubic, quartic,
and quintic order superpotential for
Model 5. The left-handed chiral superfields
are defined by Table 2.2. Note that the particle
identification for down quarks and leptons is
somewhat arbitrary: we have made a specific choice
for purposes of illustration.
\smallskip
\leftline{TABLE 5:}
\noindent The complete cubic
order superpotential for
Model 4. The left-handed chiral superfields
are defined by Table 2.1. Note that the particle
identification is somewhat arbitrary; the cubic
couplings suggest, in fact, that we should interchange
the labelling of $L_2$ and $h_1$ in order to avoid
interpreting the second term in $W_3$ as rapid proton decay.
\smallskip
\leftline{TABLES A.1-A.7:}
\noindent The basis vectors and $k_{ij}$ matrices
which define the 7 conformal field theory solutions.
The notation is explained in the appendix.
\smallskip
\leftline{TABLE A.8:}
\noindent A set of basis vectors and $k_{ij}$'s
which give a fermionic realization of
Model 5 equivalent to that obtained from
the basis vectors in Table A.5.

\vfil\eject
\centerline{TABLE 1}
\vskip .5in
{\openup2\jot
\halign{#\hfil&\quad\hfil#\hfil&\quad\hfil#\hfil&\quad\hfil#\hfil&
        \quad\hfil#\hfil&\quad$#$\hfil&\quad\hfil#\hfil\cr
 &$c_m$&$n$&$n_{trip}$&$n_{doub}$&\qquad G_{hidden}&hidden chiral\cr
\noalign{\vskip -.1in}
 & & & & & &matter?\cr
\noalign{\smallskip}
Model 1:&2&7&5&14&SU(2)_2$$\times$$[SU(2)]^2$$\times$$[U(1)]^5&no\cr
Model 2:&2&6&8&4&SO(7)$$\times$$[SU(2)_2]^2$$\times$$[SU(2)]^4&yes\cr
Model 3:&2&6&4&8&[SO(5)]^2$$\times$$SU(2)_2$$\times$$[U(1)]^4&yes\cr
Model 4:&8&5&2&12&SU(4)_2$$\times$$SU(2)_2&yes\cr
Model 5:&8&5&1&6&SU(4)_2$$\times$$SU(2)_2&yes\cr
Model 6:&6&6&3&6&[SU(2)_2]^2$$\times$$[U(1)]^3&no\cr
Model 7:&6&7&6&14&SO(7)$$\times$$SU(2)_2&yes\cr
}}
\vfil\eject
{
\catcode`@=11
\newdimen\pagewidth \newdimen\pageheight
\pagewidth=\hsize  \pageheight=\vsize
\def\onepageout#1{\shipout\vbox{
  \vbox to \pageheight{#1}
} }
\newbox\partialpage
\def\begindoublecolumns{\begingroup
  \output={\global\setbox\partialpage=\vbox{\unvbox255\bigskip}}\eject
  \output={\doublecolumnout} \hsize=2.65in \vsize=14in \parindent=0pt}
\def\enddoublecolumns{\output={\balancecolumns}\eject
  \endgroup \pagegoal=\vsize \parindent=20pt}
\def\doublecolumnout{\splittopskip=\topskip \splitmaxdepth=\maxdepth
  \dimen@=7in \advance\dimen@ by-\ht\partialpage
  \setbox0=\vsplit255 to\dimen@ \setbox2=\vsplit255 to\dimen@
  \onepageout\pagesofar \unvbox255 \penalty\outputpenalty}
\def\pagesofar{\unvbox\partialpage
  \wd0=\hsize \wd2=\hsize \hbox to\pagewidth{\box0\hfil\box2}}
\def\balancecolumns{\setbox0=\vbox{\unvbox255} \dimen@=\ht0
  \advance\dimen@ by\topskip \advance\dimen@ by-\baselineskip
  \divide\dimen@ by2 \splittopskip=\topskip
 {\vbadness=10000 \loop \global\setbox3=\copy0
    \global\setbox1=\vsplit3 to\dimen@
    \ifdim\ht3>\dimen@ \global\advance\dimen@ by1pt \repeat}
  \setbox0=\vbox to\dimen@{\unvbox1} \setbox2=\vbox to\dimen@{\unvbox3}
  \pagesofar}

\centerline{TABLE 2.1}
\vskip -.1in
\ifx\answ\bigans\begindoublecolumns\else\fi
\obeylines\baselineskip=11pt\ninepoint
Quark doublets: $(3,2)_{1/3}$
under $SU(3)_c$$\times$$SU(2)_L$$\times$$U(1)_Y$
\smallskip
$Q_1$: ( 2 8 0 -8 -28 -8 -24 )
$Q_2$: ( -2 8 -8 16 -4 -8 -24 )
$Q_3$: ( -4 8 8 -8 8 -8 -24 )
\smallskip
Up-type quark conjugates: $(\bar 3,1)_{-4/3}$
under $SU(3)_c$$\times$$SU(2)_L$$\times$$U(1)_Y$
\smallskip
$u^c_1$: ( 6 0 -8 -16 4 -64 -24 )
$u^c_2$: ( 4 16 0 0 24 -144 16 )
$u^c_3$: ( -2 -32 12 -4 16 -64 -24 )
\smallskip
Down-type quark conjugates: $(\bar 3,1)_{2/3}$
under $SU(3)_c$$\times$$SU(2)_L$$\times$$U(1)_Y$
\smallskip
$d^c_1$: ( 0 8 6 10 14 40 8 )
$d^c_2$: ( 0 8 6 10 14 40 8 )
$d^c_3$: ( 0 0 0 8 -20 32 12 )
\smallskip
Lepton doublets: $(1,2)_{-1}$
under $SU(3)_c$$\times$$SU(2)_L$$\times$$U(1)_Y$
\smallskip
$L_1$: ( 0 24 8 -8 32 -152 -8 )
$L_2$: ( -2 -16 -6 -2 14 -32 16 )
$L_3$: ( 4 -8 -8 0 12 -24 12 )
\smallskip
Lepton conjugates: $(1,1)_{2}$
under $SU(3)_c$$\times$$SU(2)_L$$\times$$U(1)_Y$
\smallskip
$e^c_1$: ( 2 48 20 20 -8 -16 8 )
$e^c_2$: ( 0 48 -4 -12 -36 48 -24 )
$e^c_3$: ( -2 32 20 4 8 64 -32 )
\smallskip
Up-type Higgs doublets: $(1,2)_{1}$
under $SU(3)_c$$\times$$SU(2)_L$$\times$$U(1)_Y$
\smallskip
$\bar h_1$: ( 0 -24 -8 8 -32 152 8 )
$\bar h_2$: ( -2 -24 8 -16 -20 152 8 )
$\bar h_3$: ( -6 -24 0 8 4 152 8 )
\smallskip
Down-type Higgs doublets: $(1,2)_{-1}$
under $SU(3)_c$$\times$$SU(2)_L$$\times$$U(1)_Y$
\smallskip
$h_1$: ( 4 -8 12 4 -28 -152 -8 )
$h_2$: ( 4 -8 -8 0 12 -24 12 )
$h_3$: ( -2 -16 -6 -2 14 -32 16 )
\smallskip
A vectorlike pair of color triplet exotics:
$(3,1)_{4/3}$$+$$(\bar 3,1)_{-4/3}$
under $SU(3)_c$$\times$$SU(2)_L$$\times$$U(1)_Y$
\smallskip
$t_1$: ( 2 0 4 4 8 160 -24 )
$\bar t_1$: ( -4 -16 -20 4 -4 -80 -16 )
\smallskip
A vectorlike pair of color triplet exotics:
$(3,1)_{-2/3}$$+$$(\bar 3,1)_{2/3}$
under $SU(3)_c$$\times$$SU(2)_L$$\times$$U(1)_Y$
\smallskip
$t_2$: ( 2 -32 0 -24 12 96 8 )
$\bar t_2$: ( 0 0 0 8 -20 32 12 )
\smallskip
Weak doublets with fractional electric charge:
$(1,2)_{0}$ under $SU(3)_c$$\times$$SU(2)_L$$\times$$U(1)_Y$
\smallskip
$D_1$: ( 4 -8 12 -4 -8 8 -4 )
$D_2$: ( 4 -8 12 -4 -8 8 -4 )
$D_3$: ( 2 16 -14 6 6 0 0 )
$D_4$: ( 2 16 -14 6 6 0 0 )
$D_5$: ( 0 16 2 -18 18 0 0 )
$D_6$: ( 0 16 2 -18 18 0 0 )
$D_7$: ( 0 -8 4 20 16 8 -4 )
$D_8$: ( 0 -8 4 20 16 8 -4 )
\smallskip
Exotics with fractional electric charge:
\smallskip
\quad$(6,2)_{1/2}$ under $SU(4)_2$$\times$$SU(2)_2$$\times$$U(1)_Y$
$T_1$: ( 0 0 10 -2 -10 16 -8 )
$T_2$: ( 0 0 10 -2 -10 16 -8 )
\smallskip
\quad$(\bar 4,2)_{-3/4}$ under $SU(4)_2$$\times$$SU(2)_2$$\times$$U(1)_Y$
$E_1$: ( 0 4 -2 -14 2 -68 -8 )
$E_2$: ( 0 4 -2 -14 2 -68 -8 )
\smallskip
\quad$(6,1)_{-1/2}$ under $SU(4)_2$$\times$$SU(2)_2$$\times$$U(1)_Y$
$S_1$: ( 0 -16 -2 2 22 32 -16 )
$S_2$: ( 0 -16 -2 2 22 32 -16 )
$S_3$: ( 0 -24 -8 0 -12 24 -12 )
$S_4$: ( 0 -24 -8 0 -12 24 -12 )
\smallskip
\quad Three vectorlike pairs of $(4,1)_{-5/4}$$+$$(\bar 4,1)_{5/4}$
\quad under $SU(4)_2$$\times$$SU(2)_2$$\times$$U(1)_Y$
$f_1$: ( 4 -12 4 4 20 -68 -8 )
$\bar f_1$: ( -4 12 -4 -4 -20 68 8 )
$f_2$: ( -2 -28 0 0 -12 -84 0 )
$\bar f_2$: ( 2 28 0 0 12 84 0 )
$f_3$: ( 0 4 -12 -12 12 -84 0 )
$\bar f_3$: ( 0 -4 12 12 -12 84 0 )
\smallskip
\quad$(4,1)_{-1/4}$ under $SU(4)_2$$\times$$SU(2)_2$$\times$$U(1)_Y$
$F_1$: ( 2 12 -2 10 14 -44 -20 )
$F_2$: ( 2 12 -2 10 14 -44 -20 )
$F_3$: ( 2 12 -2 10 14 -44 -20 )
$F_4$: ( 2 12 -2 10 14 -44 -20 )
$F_5$: ( 2 4 -8 8 -20 -52 -16 )
$F_6$: ( 0 4 8 -16 -8 -52 -16 )
$F_7$: ( -4 4 0 8 16 -52 -16 )
\smallskip
\quad$(4,1)_{3/4}$ under $SU(4)_2$$\times$$SU(2)_2$$\times$$U(1)_Y$
$F_8$: ( 2 4 -8 0 0 108 -12 )
$F_9$: ( 2 4 -8 0 0 108 -12 )
$F_{10}$: ( -4 -4 -6 -2 2 100 -8 )
$F_{11}$: ( -4 -4 -6 -2 2 100 -8 )
\smallskip
\quad$(\bar 4,1)_{1/4}$ under $SU(4)_2$$\times$$SU(2)_2$$\times$$U(1)_Y$
$\bar F_1$: ( 4 -4 0 -8 -16 52 16 )
$\bar F_2$: ( 2 28 0 8 -8 -76 -4 )
$\bar F_3$: ( 2 28 0 8 -8 -76 -4 )
$\bar F_4$: ( 0 -4 -8 16 8 52 16 )
$\bar F_5$: ( -2 -4 8 -8 20 52 16 )
$\bar F_6$: ( -4 20 2 6 -6 -84 0 )
$\bar F_7$: ( -4 20 2 6 -6 -84 0 )
\smallskip
\quad$(1,1)_{1}$ under $SU(4)_2$$\times$$SU(2)_2$$\times$$U(1)_Y$
$H_1$: ( 4 32 -8 8 -8 32 -16 )
$H_2$: ( 2 32 8 -16 4 32 -16 )
$H_3$: ( 2 0 4 20 -32 32 -16 )
$H_4$: ( 0 0 20 -4 -20 32 -16 )
$H_5$: ( -2 32 0 8 28 32 -16 )
$H_6$: ( -2 16 -12 4 -40 16 -8 )
$H_7$: ( -4 16 4 -20 -28 16 -8 )
$H_8$: ( -4 0 12 20 4 32 -16 )
$H_9$: ( -4 -8 6 18 -30 24 -12 )
$H_{10}$: ( -4 -8 6 18 -30 24 -12 )
$H_{11}$: ( -4 -8 6 18 -30 24 -12 )
$H_{12}$: ( -4 -8 6 18 -30 24 -12 )
$H_{13}$: ( -8 16 -4 4 -4 16 -8 )
\smallskip
\quad$(1,1)_{-1}$ under $SU(4)_2$$\times$$SU(2)_2$$\times$$U(1)_Y$
$\bar H_1$: ( 8 -16 4 -4 4 -16 8 )
$\bar H_2$: ( 4 0 -12 -20 -4 -32 16 )
$\bar H_3$: ( 4 -16 -4 20 28 -16 8 )
$\bar H_4$: ( 2 -16 12 -4 40 -16 8 )
$\bar H_5$: ( 2 -32 0 -8 -28 -32 16 )
$\bar H_6$: ( 0 8 6 10 14 -152 -8 )
$\bar H_7$: ( 0 8 6 10 14 -152 -8 )
$\bar H_8$: ( 0 0 0 8 -20 -160 -4 )
$\bar H_9$: ( 0 0 0 8 -20 -160 -4 )
$\bar H_{10}$: ( 0 0 -20 4 20 -32 16 )
$\bar H_{11}$: ( -2 0 -4 -20 32 -32 16 )
$\bar H_{12}$: ( -2 -32 -8 16 -4 -32 16 )
$\bar H_{13}$: ( -4 -32 8 -8 8 -32 16 )
\smallskip
Vectorlike pair of exotic singlets with electric charge $\pm$$1$:
\smallskip
$s_1$: ( -6 16 0 24 -36 48 -24 )
$\bar s_1$: ( 4 -32 -16 -16 40 32 -16 )
\smallskip
\quad$(1,3)_{0}$ under $SU(4)_2$$\times$$SU(2)_2$$\times$$U(1)_Y$
\smallskip
$T_1$: ( 6 16 4 4 32 16 -8 )
$T_2$: ( 2 0 -16 24 -12 0 0 )
$T_3$: ( -4 0 -8 24 24 0 0 )
\smallskip
\quad$(1,2)_{0}$ under $SU(4)_2$$\times$$SU(2)_2$$\times$$U(1)_Y$
\smallskip
$\varphi _1$: ( 0 -24 2 -2 -22 40 -20 )
$\varphi _2$: ( 0 -24 2 -2 -22 40 -20 )
$\varphi _3$: ( 0 -24 2 -2 -22 40 -20 )
$\varphi _4$: ( 0 -24 2 -2 -22 40 -20 )
$\varphi _5$: ( -4 0 -8 -16 4 32 -16 )
$\varphi _6$: ( -4 0 -8 -16 4 32 -16 )
$\varphi _7$: ( -4 0 -8 -16 4 32 -16 )
$\varphi _8$: ( -4 0 -8 -16 4 32 -16 )
\smallskip
Singlets with zero hypercharge and zero anomalous charge:
\smallskip
$\phi _1$: ( 6 0 -8 0 -36 0 0 )
$\phi _2$: ( 4 0 8 -24 -24 0 0 )
$\phi _3$: ( 2 32 -12 -12 24 0 0 )
$\phi _4$: ( -2 32 -20 12 48 0 0 )
$\phi _5$: ( -2 0 16 -24 12 0 0 )
$\phi _6$: ( -2 -32 12 12 -24 0 0 )
$\phi _7$: ( -4 32 -4 -12 60 0 0 )
$\phi _8$: ( -6 0 8 0 36 0 0 )
\smallskip
Singlets with zero hypercharge and nonzero anomalous charge:
\smallskip
$\chi_1$: ( 10 -16 8 16 -28 16 -8 )
$\chi_2$: ( 6 -16 0 40 -4 16 -8 )
$\chi_3$: ( 4 8 14 -22 10 8 -4 )
$\chi_4$: ( 4 8 14 -22 10 8 -4 )
$\chi_5$: ( 4 8 14 -22 10 8 -4 )
$\chi_6$: ( 4 8 14 -22 10 8 -4 )
$\chi_7$: ( 4 -16 16 16 8 16 -8 )
$\chi_8$: ( 2 16 16 -8 -4 -80 40 )
$\chi_9$: ( -2 -16 4 -28 -56 -16 8 )
$\chi_{10}$: ( -4 16 -16 -16 -8 -16 8 )
$\chi_{11}$: ( -6 -16 -4 -4 -32 -16 8 )
$\chi_{12}$: ( -8 -16 12 -28 -20 -16 8 )

\ifx\answ\bigans
\enddoublecolumns
\else
\fi

\vfil\eject

%\bye
%\input harvmac.tex
\catcode`@=11
\newdimen\pagewidth \newdimen\pageheight
\pagewidth=\hsize  \pageheight=\vsize
\def\onepageout#1{\shipout\vbox{
  \vbox to \pageheight{#1}
} }
\newbox\partialpage
\def\begindoublecolumns{\begingroup
  \output={\global\setbox\partialpage=\vbox{\unvbox255\bigskip}}\eject
  \output={\doublecolumnout} \hsize=2.65in \vsize=14in \parindent=0pt}
\def\enddoublecolumns{\output={\balancecolumns}\eject
  \endgroup \pagegoal=\vsize \parindent=20pt}
\def\doublecolumnout{\splittopskip=\topskip \splitmaxdepth=\maxdepth
  \dimen@=7in \advance\dimen@ by-\ht\partialpage
  \setbox0=\vsplit255 to\dimen@ \setbox2=\vsplit255 to\dimen@
  \onepageout\pagesofar \unvbox255 \penalty\outputpenalty}
\def\pagesofar{\unvbox\partialpage
  \wd0=\hsize \wd2=\hsize \hbox to\pagewidth{\box0\hfil\box2}}
\def\balancecolumns{\setbox0=\vbox{\unvbox255} \dimen@=\ht0
  \advance\dimen@ by\topskip \advance\dimen@ by-\baselineskip
  \divide\dimen@ by2 \splittopskip=\topskip
 {\vbadness=10000 \loop \global\setbox3=\copy0
    \global\setbox1=\vsplit3 to\dimen@
    \ifdim\ht3>\dimen@ \global\advance\dimen@ by1pt \repeat}
  \setbox0=\vbox to\dimen@{\unvbox1} \setbox2=\vbox to\dimen@{\unvbox3}
  \pagesofar}

\centerline{TABLE 2.2}
\ifx\answ\bigans\begindoublecolumns\else\fi
\obeylines\baselineskip=12pt\ninepoint
Quark doublets: $(3,2)_{1/3}$
under $SU(3)_c$$\times$$SU(2)_L$$\times$$U(1)_Y$
\smallskip
$Q_1$: ( -2 0 8 -2 -8 16 -16 )
$Q_2$: ( 2 0 8 -2 -8 16 -16 )
$Q_3$: ( 0 -2 0 -2 20 4 -12 )
\smallskip
Up-type quark conjugates: $(\bar 3,1)_{-4/3}$
under $SU(3)_c$$\times$$SU(2)_L$$\times$$U(1)_Y$
\smallskip
$u^c_1$: ( 2 0 0 6 0 -16 -16 ) -4
$u^c_2$: ( -2 0 0 6 0 -16 -16 ) -4
$u^c_3$: ( 0 6 -8 2 -4 -28 -12 ) -4
\smallskip
Down-type quark conjugates: $(\bar 3,1)_{2/3}$
under $SU(3)_c$$\times$$SU(2)_L$$\times$$U(1)_Y$
\smallskip
$d^c_1$: ( 0 -6 -8 2 -4 -28 -12 )
$d^c_2$: ( 0 0 0 0 -12 20 4 )
$d^c_3$: ( 0 2 -8 -2 0 8 8 )
\smallskip
Lepton doublets: $(1,2)_{-1}$
under $SU(3)_c$$\times$$SU(2)_L$$\times$$U(1)_Y$
\smallskip
$L_1$: ( 0 4 -8 0 16 -24 -24 )
$L_2$: ( 0 2 16 -2 -12 -12 -28 )
$L_3$: ( 0 2 0 2 16 0 0 )
\smallskip
Lepton conjugates: $(1,1)_{2}$
under $SU(3)_c$$\times$$SU(2)_L$$\times$$U(1)_Y$
\smallskip
$e^c_1$: ( 0 -2 -8 -6 4 12 -4 )
$e^c_2$: ( 4 -2 -24 -2 -4 -12 4 )
$e^c_3$: ( 0 -2 -8 -6 4 12 -4 )
\smallskip
Up-type Higgs doublets: $(1,2)_{1}$
under $SU(3)_c$$\times$$SU(2)_L$$\times$$U(1)_Y$
\smallskip
$\bar h_1$: ( 0 0 -8 -4 -4 -12 4 )
$\bar h_2$: ( 0 -2 -16 2 0 0 0 )
$\bar h_3$: ( 0 -4 8 0 -16 24 24 )
$\bar h_4$: ( 0 -2 0 -2 8 24 -8 )
\smallskip
Down-type Higgs doublets: $(1,2)_{-1}$
under $SU(3)_c$$\times$$SU(2)_L$$\times$$U(1)_Y$
\smallskip
$h_1$: ( 0 0 8 4 4 12 -4 )
$h_2$: ( 0 2 0 2 16 0 0 )
$h_3$: ( 0 2 0 2 -8 -24 8 )
$h_4$: ( 0 2 16 -2 0 0 0 )
\smallskip
Vectorlike pair of color triplet
exotics: $(3,1)_{-2/3}$$+$$(\bar 3,1)_{2/3}$
under $SU(3)_c$$\times$$SU(2)_L$$\times$$U(1)_Y$
\smallskip
$t$: ( 0 0 0 0 12 -20 -4 )
$\bar t$: ( 0 2 -8 -2 0 8 8 )
\smallskip
Weak doublet exotics with fractional electric charge:
$(2,2)_{0}$ under $SU(2)_L$$\times$$SU(2)_2$$\times$$U(1)_Y$
\smallskip
$D_1$: ( 0 0 -8 2 -4 -12 4 )
$D_2$: ( 0 0 -8 2 -4 -12 4 )
$D_3$: ( 0 0 -8 2 -4 -12 4 )
$D_4$: ( 0 0 -8 2 -4 -12 4 )
\smallskip
Exotics with fractional electric charge:
\smallskip
\quad$(4,1)_{-1}$ under $SU(4)_2$$\times$$SU(2)_2$$\times$$U(1)_Y$
$F_1$: ( 0 0 12 0 12 0 -16 )
$F_2$: ( 0 0 12 0 12 0 -16 )
\smallskip
\quad$(\bar 4,1)_{1}$ under $SU(4)_2$$\times$$SU(2)_2$$\times$$U(1)_Y$
$\bar F_1$: ( 0 0 4 -4 -4 24 8 )
$\bar F_2$: ( 0 0 4 -4 -4 24 8 )
$\bar F_3$: ( 0 0 -12 0 0 12 -20 )
$\bar F_4$: ( 0 0 -12 0 0 12 -20 )
$\bar F_5$: ( 0 0 -12 0 0 12 -20 )
$\bar F_6$: ( 0 0 -12 0 0 12 -20 )
\smallskip
\quad$(\bar 4,1)_{-1}$ under $SU(4)_2$$\times$$SU(2)_2$$\times$$U(1)_Y$
$\bar F_7$:  ( 2 0 4 2 8 -12 -12 )
$\bar F_8$:  ( 2 0 4 2 8 -12 -12 )
$\bar F_9$:  ( -2 0 4 2 8 -12 -12 )
$\bar F_{10}$: ( -2 0 4 2 8 -12 -12 )
\smallskip
\quad$(1,2)_{1}$ under $SU(4)_2$$\times$$SU(2)_2$$\times$$U(1)_Y$
$H_1$: ( 0 -2 8 -4 -12 12 -4 )
$H_2$: ( 0 -2 8 -4 -12 12 -4 )
$H_3$: ( 0 -4 0 2 -8 24 -8 )
$H_4$: ( 0 -4 0 2 -8 24 -8 )
\smallskip
\quad$(1,2)_{-1}$ under $SU(4)_2$$\times$$SU(2)_2$$\times$$U(1)_Y$
$H_5$: ( 0 2 -8 4 12 -12 4 )
$H_6$: ( 0 0 16 2 -16 0 0 )
$H_7$: ( 0 2 -8 4 12 -12 4 )
$H_8$: ( 0 0 16 2 -16 0 0 )
\smallskip
Singlet exotics with electric charge $+$$1$:
\smallskip
$s_1$: ( 0 -4 0 -4 -8 24 -8 )
$s_2$: ( 2 0 -16 -2 -32 0 0 )
$s_3$: ( 0 -4 -16 0 8 24 -8 )
$s_4$: ( 0 -4 0 -4 -8 24 -8 )
$s_5$: ( -2 0 -16 -2 -32 0 0 )
$s_6$: ( -4 -2 -24 -2 -4 -12 4 )
\smallskip
Singlet exotics with electric charge $-$$1$:
\smallskip
$\bar s_1$: ( 0 4 16 0 16 0 0 )
$\bar s_2$: ( 0 4 16 0 16 0 0 )
$\bar s_3$: ( 0 4 16 0 -8 -24 8 )
$\bar s_4$: ( 0 2 24 2 4 12 -4 )
$\bar s_5$: ( 0 2 24 2 4 12 -4 )
$\bar s_6$: ( 0 -2 8 2 12 -60 20 )
\smallskip
Singlets with zero hypercharge and zero anomalous charge:
\smallskip
$\phi _1$: ( 4 0 0 0 0 0 0 )
$\phi _2$: ( 2 4 0 -2 -16 0 0 )
$\phi _3$: ( 2 4 0 -2 -16 0 0 )
$\phi _4$: ( 2 -4 16 -2 0 0 0 )
$\phi _5$: ( 2 -4 16 -2 0 0 0 )
$\phi _6$: ( 2 -4 0 2 16 0 0 )
$\phi _7$: ( 2 -4 0 2 16 0 0 )
$\phi _8$: ( 0 0 -16 4 16 0 0 )
$\phi _9$: ( 0 0 -16 4 16 0 0 )
$\phi _{10}$: ( -2 4 0 -2 -16 0 0 )
$\phi _{11}$: ( -2 4 0 -2 -16 0 0 )
$\phi _{12}$: ( -2 -4 16 -2 0 0 0 )
$\phi _{13}$: ( -2 -4 16 -2 0 0 0 )
$\phi _{14}$: ( -2 -4 0 2 16 0 0 )
$\phi _{15}$: ( -2 -4 0 2 16 0 0 )
$\phi _{16}$: ( -4 0 0 0 0 0 0 )
\smallskip
Hidden fields:
\smallskip
\quad$(4,2)_{0}$ under $SU(4)_2$$\times$$SU(2)_2$$\times$$U(1)_Y$
$E_1$: ( 0 0 -4 -2 4 -24 -8 )
$E_2$: ( 0 0 -4 -2 4 -24 -8 )
$E_3$: ( 0 -2 4 0 4 0 16 )
$E_4$: ( 0 -2 4 0 -8 -12 -12 )
$E_5$: ( 0 -2 4 0 -8 -12 -12 )
\smallskip
\quad$(\bar 4,2)_{0}$ under $SU(4)_2$$\times$$SU(2)_2$$\times$$U(1)_Y$
$\bar E_1$: ( 0 2 -4 0 -4 0 -16 )
\smallskip
\quad$(6,1)_{0}$ under $SU(4)_2$$\times$$SU(2)_2$$\times$$U(1)_Y$
$S_1$: ( 2 0 -8 2 -4 -12 4 )
$S_2$: ( 0 -4 8 0 -4 -12 4 )
$S_3$: ( -2 0 -8 2 -4 -12 4 )
$S_4$: ( 2 0 8 -2 4 12 -4 )
$S_5$: ( 0 4 -8 0 4 12 -4 )
$S_6$: ( 0 2 0 2 -8 24 -8 )
$S_7$: ( 0 2 0 2 -8 24 -8 )
$S_8$: ( -2 0 8 -2 4 12 -4 )
\smallskip
\quad$(1,3)_{0}$ under $SU(4)_2$$\times$$SU(2)_2$$\times$$U(1)_Y$
$T_1$: ( 2 4 0 -2 8 24 -8 )
$T_2$: ( 0 -2 8 2 12 36 -12 )
$T_3$: ( -2 4 0 -2 8 24 -8 )
\smallskip
Singlets with zero hypercharge and nonzero anomalous charge:
\smallskip
$\chi_1$: ( 2 4 0 -2 8 24 -8 )
$\chi_2$: ( 0 2 8 -6 -4 -12 4 )
$\chi_3$: ( 0 2 8 -6 -4 -12 4 )
$\chi_4$: ( 0 2 -8 -2 0 -24 -24 )
$\chi_5$: ( 0 2 -8 -2 0 -24 -24 )
$\chi_6$: ( 0 2 -24 2 28 -12 4 )
$\chi_7$: ( 0 0 16 -4 8 24 -8 )
$\chi_8$: ( 0 0 0 0 -12 -12 -28 )
$\chi_9$: ( 0 -2 8 2 12 36 -12 )
$\chi_{10}$: ( -2 4 0 -2 8 24 -8 )
$\chi_{11}$: ( 0 0 0 0 12 12 28 )
$\chi_{12}$: ( 0 0 0 0 -24 -24 8 )
$\chi_{13}$: ( 0 0 0 0 -24 -24 8 )
$\chi_{14}$: ( 0 0 -16 4 -8 -24 8 )

\ifx\answ\bigans
\enddoublecolumns
\else
\fi

\vfil\eject
%\bye
%\input harvmac.tex
\catcode`@=11
\newdimen\pagewidth \newdimen\pageheight
\pagewidth=\hsize  \pageheight=\vsize
\def\onepageout#1{\shipout\vbox{
  \vbox to \pageheight{#1}
} }
\newbox\partialpage
\def\begindoublecolumns{\begingroup
  \output={\global\setbox\partialpage=\vbox{\unvbox255\bigskip}}\eject
  \output={\doublecolumnout} \hsize=2.65in \vsize=14in \parindent=0pt}
\def\enddoublecolumns{\output={\balancecolumns}\eject
  \endgroup \pagegoal=\vsize \parindent=20pt}
\def\doublecolumnout{\splittopskip=\topskip \splitmaxdepth=\maxdepth
  \dimen@=7in \advance\dimen@ by-\ht\partialpage
  \setbox0=\vsplit255 to\dimen@ \setbox2=\vsplit255 to\dimen@
  \onepageout\pagesofar \unvbox255 \penalty\outputpenalty}
\def\pagesofar{\unvbox\partialpage
  \wd0=\hsize \wd2=\hsize \hbox to\pagewidth{\box0\hfil\box2}}
\def\balancecolumns{\setbox0=\vbox{\unvbox255} \dimen@=\ht0
  \advance\dimen@ by\topskip \advance\dimen@ by-\baselineskip
  \divide\dimen@ by2 \splittopskip=\topskip
 {\vbadness=10000 \loop \global\setbox3=\copy0
    \global\setbox1=\vsplit3 to\dimen@
    \ifdim\ht3>\dimen@ \global\advance\dimen@ by1pt \repeat}
  \setbox0=\vbox to\dimen@{\unvbox1} \setbox2=\vbox to\dimen@{\unvbox3}
  \pagesofar}

\centerline{TABLE 2.3}
\vskip -.2in
\ifx\answ\bigans\begindoublecolumns\else\fi
\obeylines\baselineskip=12pt\ninepoint
Quark doublets: $(3,2)_{1/3}$
under $SU(3)_c$$\times$$SU(2)_L$$\times$$U(1)_Y$
\smallskip
$Q_1$:  ( 0 0 0 4 0 0 -8 -4 16 24 -24 )
$Q_2$:  ( 0 0 0 2 0 -4 -4 6 0 -16 -32 )
$Q_3$:  ( 0 0 0 -2 0 -4 -4 -6 0 -16 -32 )
\smallskip
Up-type quark conjugates: $(\bar 3,1)_{-4/3}$
under $SU(3)_c$$\times$$SU(2)_L$$\times$$U(1)_Y$
\smallskip
$u^c_1$:  ( 0 0 0 2 4 -8 -4 6 -8 16 -16 )
$u^c_2$:  ( 0 0 0 0 0 0 24 0 32 -32 -16 )
$u^c_3$:  ( 0 0 0 -2 4 -8 -4 -6 -8 16 -16 )
\smallskip
Down-type quark conjugates: $(\bar 3,1)_{2/3}$
under $SU(3)_c$$\times$$SU(2)_L$$\times$$U(1)_Y$
\smallskip
$d^c_1$:  ( 0 0 0 2 0 12 -4 6 -8 16 -16 )
$d^c_2$:  ( 0 0 0 -2 0 12 -4 -6 -8 16 -16 )
$d^c_3$:  ( 0 0 0 -4 0 8 0 4 -24 -24 -24 )
\smallskip
Lepton doublets: $(1,2)_{-1}$
under $SU(3)_c$$\times$$SU(2)_L$$\times$$U(1)_Y$
\smallskip
$L_1$:  ( 0 0 0 2 4 8 -4 6 -16 48 0 )
$L_2$:  ( 0 0 0 -2 4 8 -4 -6 -16 48 0 )
$L_3$:  ( 0 0 0 0 0 -8 8 0 8 0 48 )
\smallskip
Lepton conjugates: $(1,1)_{2}$
under $SU(3)_c$$\times$$SU(2)_L$$\times$$U(1)_Y$
\smallskip
$e^c_1$:  ( 0 0 0 2 -4 0 -4 6 8 -48 -48 )
$e^c_2$:  ( 0 0 0 -2 -4 0 -4 -6 8 -48 -48 )
$e^c_3$:  ( 0 0 0 -2 0 8 -8 2 12 -20 -28 )
\smallskip
Up-type Higgs doublets: $(1,2)_{1}$
under $SU(3)_c$$\times$$SU(2)_L$$\times$$U(1)_Y$
\smallskip
$\bar h_1$:  ( 0 0 0 -2 0 0 0 2 20 -20 20 )
$\bar h_2$:  ( 0 0 0 0 0 8 -8 0 -8 0 -48 )
$\bar h_3$:  ( 0 0 0 4 -4 -4 0 -4 32 -8 8 )
\smallskip
Down-type Higgs doublet: $(1,2)_{-1}$
under $SU(3)_c$$\times$$SU(2)_L$$\times$$U(1)_Y$
\smallskip
$h$:  ( 0 0 0 2 0 0 0 -2 -20 20 -20 )
\smallskip
A vectorlike pair of weak doublet exotics:
$(1,2)_{3}$$+$$(1,2)_{-3}$ under
$SU(3)_c$$\times$$SU(2)_L$$\times$$U(1)_Y$
\smallskip
$\bar h_1$:  ( 0 0 0 -4 -4 4 -16 4 0 56 -8 )
$h_1$:  ( 0 0 0 4 4 -4 16 -4 0 -56 8 )
\smallskip
Weak doublet exotics:
$(1,2)_3$ under
$SU(3)_c$$\times$$SU(2)_L$$\times$$U(1)_Y$
\smallskip
$\bar h_4$:  ( 0 0 0 0 0 8 -16 0 8 -96 0 )
$\bar h_5$:  ( 0 0 0 0 -4 4 -24 0 -24 0 0 )
\smallskip
A vectorlike pair of color triplet
exotics: $(3,1)_{-2/3}$$+$$(\bar 3,1)_{2/3}$
under $SU(3)_c$$\times$$SU(2)_L$$\times$$U(1)_Y$
\smallskip
$t_1$:  ( 0 0 0 2 0 0 8 -2 4 -44 -4 )
$\bar t_1$:  ( 0 0 0 -2 0 0 -8 2 -4 44 4 )
\smallskip
Two vectorlike pairs of color triplet
exotics: $(3,1)_{4/3}$$+$$(\bar 3,1)_{-4/3}$
under $SU(3)_c$$\times$$SU(2)_L$$\times$$U(1)_Y$
\smallskip
$t_2$:  ( 0 0 0 -4 4 4 -8 4 24 -8 8 )
$\bar t_2$:  ( 0 0 0 4 -4 -4 8 -4 -24 8 -8 )
$t_3$:  ( 0 0 0 -4 -4 4 8 4 -8 -56 8 )
$\bar t_3$:  ( 0 0 0 0 -4 -4 16 0 0 64 -16 )
\smallskip
Weak doublet exotics:
$(2,2)_{-1}$ under $SU(2)_L$$\times$$SU(2)_2^{(1)}$$\times$$U(1)_Y$
\smallskip
$D_1$:  ( 2 0 0 -2 0 0 12 2 -4 4 -4 )
$D_2$:  ( 0 0 2 -2 0 0 12 2 -4 4 -4 )
$D_3$:  ( 0 0 -2 -2 0 0 12 2 -4 4 -4 )
$D_4$:  ( -2 0 0 -2 0 0 12 2 -4 4 -4 )
\smallskip
Exotics with fractional electric charge:
$(2,1)_{1}$ or $(1,2)_{1}$ under
$SU(2)_2^{(1)}$$\times$$SU(2)_2^{(2)}$$\times$$U(1)_Y$
\smallskip
$H_1^{(1)}$:  ( 2 2 0 0 -2 6 0 0 12 12 -12 )
$H_2^{(1)}$:  ( 2 2 0 -2 2 2 -8 2 0 -32 -16 )
$H_3^{(2)}$:  ( 2 2 0 -2 0 0 -12 2 -16 16 -16 )
$H_4^{(1)}$:  ( 2 -2 0 0 -2 6 0 0 12 12 -12 )
$H_5^{(1)}$:  ( 2 -2 0 -2 2 2 -8 2 0 -32 -16 )
$H_6^{(2)}$:  ( 2 -2 0 -2 0 0 -12 2 -16 16 -16 )
$H_7^{(1)}$:  ( 0 2 2 0 -2 6 0 0 12 12 -12 )
$H_8^{(1)}$:  ( 0 2 2 -2 2 2 -8 2 0 -32 -16 )
$H_9^{(2)}$:  ( 0 2 2 -2 0 0 -12 2 -16 16 -16 )
$H_{10}^{(1)}$:  ( 0 2 -2 0 -2 6 0 0 12 12 -12 )
$H_{11}^{(1)}$:  ( 0 2 -2 -2 2 2 -8 2 0 -32 -16 )
$H_{12}^{(2)}$:  ( 0 2 -2 -2 0 0 -12 2 -16 16 -16 )
$H_{13}^{(1)}$:  ( 0 -2 2 0 -2 6 0 0 12 12 -12 )
$H_{14}^{(1)}$:  ( 0 -2 2 -2 2 2 -8 2 0 -32 -16 )
$H_{15}^{(2)}$:  ( 0 -2 2 -2 0 0 -12 2 -16 16 -16 )
$H_{16}^{(1)}$:  ( 0 -2 -2 0 -2 6 0 0 12 12 -12 )
$H_{17}^{(1)}$:  ( 0 -2 -2 -2 2 2 -8 2 0 -32 -16 )
$H_{18}^{(2)}$:  ( 0 -2 -2 -2 0 0 -12 2 -16 16 -16 )
$H_{19}^{(1)}$:  ( -2 2 0 0 -2 6 0 0 12 12 -12 )
$H_{20}^{(1)}$:  ( -2 2 0 -2 2 2 -8 2 0 -32 -16 )
$H_{21}^{(2)}$:  ( -2 2 0 -2 0 0 -12 2 -16 16 -16 )
$H_{22}^{(1)}$:  ( -2 -2 0 0 -2 6 0 0 12 12 -12 )
$H_{23}^{(1)}$:  ( -2 -2 0 -2 2 2 -8 2 0 -32 -16 )
$H_{24}^{(2)}$:  ( -2 -2 0 -2 0 0 -12 2 -16 16 -16 )
\smallskip
Exotics with unit electric charge:
\smallskip
\quad$(1,2)_2$ under
$SU(2)_2^{(1)}$$\times$$SU(2)_2^{(2)}$$\times$$U(1)_Y$
$\delta _1$:  ( 2 0 0 -2 -2 -2 -8 2 12 -20 -28 )
$\delta _2$:  ( 0 0 2 -2 -2 -2 -8 2 12 -20 -28 )
$\delta _3$:  ( 0 0 -2 -2 -2 -2 -8 2 12 -20 -28 )
$\delta _4$:  ( -2 0 0 -2 -2 -2 -8 2 12 -20 -28 )
\smallskip
\quad$(2,1)_{-2}$ or $(1,2)_{-2}$ under
$SU(2)_2^{(1)}$$\times$$SU(2)_2^{(2)}$$\times$$U(1)_Y$
$\bar \delta _1^{(2)}$:  ( 2 0 0 2 2 -6 8 -2 12 44 4 )
$\bar \delta _2^{(2)}$:  ( 2 0 0 0 2 2 16 0 -8 -24 -24 )
$\bar \delta _3^{(1)}$:  ( 2 0 0 0 0 0 12 0 -24 24 -24 )
$\bar \delta _4^{(2)}$:  ( 0 0 2 2 2 -6 8 -2 12 44 4 )
$\bar \delta _5^{(2)}$:  ( 0 0 2 0 2 2 16 0 -8 -24 -24 )
$\bar \delta _6^{(1)}$:  ( 0 0 2 0 0 0 12 0 -24 24 -24 )
$\bar \delta _7^{(2)}$:  ( 0 0 -2 2 2 -6 8 -2 12 44 4 )
$\bar \delta _8^{(2)}$:  ( 0 0 -2 0 2 2 16 0 -8 -24 -24 )
$\bar \delta _9^{(1)}$:  ( 0 0 -2 0 0 0 12 0 -24 24 -24 )
$\bar \delta _{10}^{(2)}$:  ( -2 0 0 2 2 -6 8 -2 12 44 4 )
$\bar \delta _{11}^{(2)}$:  ( -2 0 0 0 2 2 16 0 -8 -24 -24 )
$\bar \delta _{12}^{(1)}$:  ( -2 0 0 0 0 0 12 0 -24 24 -24 )
\smallskip
Singlets with electric charge $+$$2$
\smallskip
$\tau_1$:  ( 0 0 0 2 0 -4 -36 6 24 0 0 )
$\tau_2$:  ( 0 0 0 -2 0 -4 -36 -6 24 0 0 )
\smallskip
Singlet with electric charge $-$$2$
\smallskip
$\bar \tau_1$:  ( 0 0 0 4 0 8 32 -4 -8 40 8 )
\smallskip
Singlets with electric charge $+$$1$
\smallskip
$s_1$:  ( 0 0 0 4 -4 0 -12 4 4 -4 4 )
$s_2$:  ( 0 0 0 0 0 8 -16 0 8 24 24 )
$s_3$:  ( 0 0 0 0 -4 0 -12 -8 4 -4 4 )
$s_4$:  ( 0 0 0 -2 0 4 -12 10 -8 -16 16 )
$s_5$:  ( 0 0 0 -2 -4 -4 -8 2 -12 -44 -4 )
$s_6$:  ( 0 0 0 -4 -4 -4 -8 4 8 56 40 )
$s_7$:  ( 0 0 0 -6 0 4 -12 -2 -8 -16 16 )
\smallskip
Singlets with electric charge $-$$1$
\smallskip
$\bar s_1$:  ( 0 4 0 6 0 -4 12 2 8 16 -16 )
$\bar s_2$:  ( 0 4 0 2 0 -4 12 -10 8 16 -16 )
$\bar s_3$:  ( 0 4 0 0 0 -8 16 0 -8 -24 -24 )
$\bar s_4$:  ( 0 0 0 6 0 -4 12 2 8 16 -16 )
$\bar s_5$:  ( 0 0 0 4 4 4 8 -4 -8 -56 -40 )
$\bar s_6$:  ( 0 0 0 2 4 4 8 -2 12 44 4 )
$\bar s_7$:  ( 0 0 0 2 0 -4 12 -10 8 16 -16 )
$\bar s_8$:  ( 0 0 0 2 0 -8 8 -2 -12 20 28 )
$\bar s_9$:  ( 0 0 0 0 4 4 16 0 16 0 -48 )
$\bar s_{10}$:  ( 0 0 0 0 4 0 12 8 -4 4 -4 )
$\bar s_{11}$:  ( 0 0 0 0 0 0 8 0 -16 96 -48 )
$\bar s_{12}$:  ( 0 0 0 0 0 -8 16 0 -8 -24 -24 )
$\bar s_{13}$:  ( 0 0 0 -4 4 0 12 -4 -4 4 -4 )
$\bar s_{14}$:  ( 0 0 0 -4 4 -4 8 4 -24 8 40 )
$\bar s_{15}$:  ( 0 -4 0 6 0 -4 12 2 8 16 -16 )
$\bar s_{16}$:  ( 0 -4 0 2 0 -4 12 -10 8 16 -16 )
$\bar s_{17}$:  ( 0 -4 0 0 0 -8 16 0 -8 -24 -24 )
\smallskip
Hidden fields:
\smallskip
\quad$(3,3)$ under $SU(2)_2^{(1)}$$\times$$SU(2)_2^{(2)}$
$N$:  ( 0 0 0 0 0 0 0 0 0 0 0 )
\smallskip
\quad$(3,1)$ and $(1,3)$, resp., under
$SU(2)_2^{(1)}$$\times$$SU(2)_2^{(2)}$
$T_1$:  ( 0 0 0 4 0 -8 -8 -4 0 -32 -16 )
$T_2$:  ( 0 0 0 -4 0 8 8 4 0 32 16 )
\quad$(2,1)$ or $(1,2)$ under
$SU(2)_2^{(1)}$$\times$$SU(2)_2^{(2)}$
$\varphi _1^{(2)}$:  ( 2 0 0 4 2 2 -8 -4 0 -32 -16 )
$\varphi _2^{(1)}$:  ( 2 0 0 4 0 0 -12 -4 -16 16 -16 )
$\varphi _3^{(2)}$:  ( 2 0 0 4 -2 2 4 4 -4 -28 -20 )
$\varphi _4^{(2)}$:  ( 2 0 0 0 2 2 0 0 24 24 -24 )
$\varphi _5^{(1)}$:  ( 2 0 0 0 0 0 -4 0 8 72 -24 )
$\varphi _6^{(1)}$:  ( 2 0 0 0 0 -8 4 0 16 -48 0 )
$\varphi _7^{(2)}$:  ( 2 0 0 0 -2 2 4 -8 -4 -28 -20 )
$\varphi _8^{(2)}$:  ( 2 0 0 0 -2 -10 0 0 0 0 0 )
$\varphi _9^{(2)}$:  ( 0 0 2 4 2 2 -8 -4 0 -32 -16 )
$\varphi _{10}^{(1)}$:  ( 0 0 2 4 0 0 -12 -4 -16 16 -16 )
$\varphi _{11}^{(2)}$:  ( 0 0 2 4 -2 2 4 4 -4 -28 -20 )
$\varphi _{12}^{(2)}$:  ( 0 0 2 0 2 2 0 0 24 24 -24 )
$\varphi _{13}^{(1)}$:  ( 0 0 2 0 0 0 -4 0 8 72 -24 )
$\varphi _{14}^{(1)}$:  ( 0 0 2 0 0 -8 4 0 16 -48 0 )
$\varphi _{15}^{(2)}$:  ( 0 0 2 0 -2 2 4 -8 -4 -28 -20 )
$\varphi _{16}^{(2)}$:  ( 0 0 2 0 -2 -10 0 0 0 0 0 )
$\varphi _{17}^{(2)}$:  ( 0 0 -2 4 2 2 -8 -4 0 -32 -16 )
$\varphi _{18}^{(1)}$:  ( 0 0 -2 4 0 0 -12 -4 -16 16 -16 )
$\varphi _{19}^{(2)}$:  ( 0 0 -2 4 -2 2 4 4 -4 -28 -20 )
$\varphi _{20}^{(2)}$:  ( 0 0 -2 0 2 2 0 0 24 24 -24 )
$\varphi _{21}^{(1)}$:  ( 0 0 -2 0 0 0 -4 0 8 72 -24 )
$\varphi _{22}^{(1)}$:  ( 0 0 -2 0 0 -8 4 0 16 -48 0 )
$\varphi _{23}^{(2)}$:  ( 0 0 -2 0 -2 2 4 -8 -4 -28 -20 )
$\varphi _{24}^{(2)}$:  ( 0 0 -2 0 -2 -10 0 0 0 0 0 )
$\varphi _{25}^{(2)}$:  ( -2 0 0 4 2 2 -8 -4 0 -32 -16 )
$\varphi _{26}^{(1)}$:  ( -2 0 0 4 0 0 -12 -4 -16 16 -16 )
$\varphi _{27}^{(2)}$:  ( -2 0 0 4 -2 2 4 4 -4 -28 -20 )
$\varphi _{28}^{(2)}$:  ( -2 0 0 0 2 2 0 0 24 24 -24 )
$\varphi _{29}^{(1)}$:  ( -2 0 0 0 0 0 -4 0 8 72 -24 )
$\varphi _{30}^{(1)}$:  ( -2 0 0 0 0 -8 4 0 16 -48 0 )
$\varphi _{31}^{(2)}$:  ( -2 0 0 0 -2 2 4 -8 -4 -28 -20 )
$\varphi _{32}^{(2)}$:  ( -2 0 0 0 -2 -10 0 0 0 0 0 )
\smallskip
Singlets with zero hypercharge and zero
anomalous charge:
\smallskip
$\phi _1$:  ( 4 0 4 0 0 0 0 0 0 0 0 )
$\phi _2$:  ( 4 0 -4 0 0 0 0 0 0 0 0 )
$\phi _3$:  ( 0 0 0 2 0 -4 -4 6 -40 -96 0 )
$\phi _4$:  ( 0 0 0 0 0 0 0 0 0 0 0 )
$\phi _5$:  ( 0 0 0 0 0 0 0 0 0 0 0 )
$\phi _6$:  ( 0 0 0 0 0 0 0 0 0 0 0 )
$\phi _7$:  ( 0 0 0 0 0 0 0 0 0 0 0 )
$\phi _8$:  ( 0 0 0 0 0 0 0 0 0 0 0 )
$\phi _9$:  ( 0 0 0 -2 0 -4 -4 -6 -40 -96 0 )
$\phi _{10}$:  ( -4 0 4 0 0 0 0 0 0 0 0 )
$\phi _{11}$:  ( -4 0 -4 0 0 0 0 0 0 0 0 )
\smallskip
Singlets with zero hypercharge and nonzero
anomalous charge:
\smallskip
$\chi_1$:  ( 0 0 0 4 4 0 -12 4 4 -4 4 )
$\chi_2$:  ( 0 0 0 4 0 -8 -8 -4 0 -32 -16 )
$\chi_3$:  ( 0 0 0 2 -4 4 8 -2 12 44 4 )
$\chi_4$:  ( 0 0 0 0 4 0 -12 -8 4 -4 4 )
$\chi_5$:  ( 0 0 0 0 0 8 0 0 -24 -24 24 )
$\chi_6$:  ( 0 0 0 0 0 8 0 0 -24 -24 24 )
$\chi_7$:  ( 0 0 0 0 0 0 8 16 24 56 -8 )
$\chi_8$:  ( 0 0 0 0 0 -8 0 0 24 24 -24 )
$\chi_9$:  ( 0 0 0 0 0 -8 0 0 24 24 -24 )
$\chi_{10}$:  ( 0 0 0 0 -4 0 12 8 -4 4 -4 )
$\chi_{11}$:  ( 0 0 0 -2 4 -4 -8 2 -12 -44 -4 )
$\chi_{12}$:  ( 0 0 0 -4 0 8 8 4 0 32 16 )
$\chi_{13}$:  ( 0 0 0 -4 -4 0 12 -4 -4 4 -4 )
$\chi_{14}$:  ( 0 0 0 -6 0 -4 12 14 8 16 -16 )
$\chi_{15}$:  ( 0 0 0 -8 0 0 8 -8 24 56 -8 )
$\chi_{16}$:  ( 0 0 0 -10 0 -4 12 2 8 16 -16 )

\ifx\answ\bigans
\enddoublecolumns
\else
\fi
}
\vfil\eject
\centerline{TABLE 3}
\vskip .5in
{\openup2\jot
\halign{#\hfil&\qquad #\hfil&\qquad\qquad\qquad#\hfil\cr
Model 1:&A,B,A'&$V_4$, $V_5$, $V_4$$+$$2V_3$$+$$2V_5$$+$$2V_6$\cr
Model 2:&A,B,A'&$V_4$, $V_5$, $V_4$$+$$2V_3$$+$$2V_5$$+$$2V_6$\cr
Model 3:&A,B,A'&$V_4$, $V_5$, $V_4$$+$$2V_3$$+$$2V_5$$+$$2V_6$\cr
Model 4:&A,B,A+B&$V_4$, $V_5$, $V_1$$+$$V_4$$+$$3V_5$\cr
Model 5:&A,A,B&$V_4$, $3V_5$, $3V_5$\cr
Model 6:&A,B,A'&$V_4$, $V_5$, $V_4$$+$$2V_3$$+$$2V_5$$+$$2V_6$\cr
Model 7:&A,A,B&$V_4$$+$$2V_3$$+$$2V_5$$+$$2V_6$, $V_5$, $V_5$\cr
}
}

\vfil\eject

{
$W_3$ = $Q_3$ $u^c_3$ $\bar h_3$ + $Q_3$ $d^c_2$ $h_3$
+ $L_2$ $\bar h_2$ $\chi_{11}$ + $L_2$ $\bar h_3$ $\chi_6$\par
 + $L_3$ $\bar h_3$ $\chi_5$ + $L_3$ $\bar h_4$ $\chi_{13}$
+ $L_3$ $h_3$ $s_4$ + $e^c_1$ $h_1$ $h_3$\par
 + $\bar h_1$ $\bar h_2$ $\bar s_4$ + $\bar h_1$ $h_3$ $\chi_9$
+ $\bar h_2$ $\bar h_4$ $\bar s_3$ + $\bar h_2$ $h_1$ $\chi_2$\par
 + $\bar h_3$ $h_2$ $\chi_4$ + $\bar h_4$ $h_2$ $\chi_{12}$
+ $h_2$ $h_3$ $s_1$ + $h_3$ $h_4$ $s_3$\par
 + $F_1$ $\bar F_1$ $\chi_{14}$ + $F_2$ $\bar F_2$ $\chi_{14}$
+ $\bar F_1$ $\bar F_7$ $S_3$ + $\bar F_1$ $\bar F_9$ $S_1$\par
 + $\bar F_1$ $H_5$ $E_5$ + $\bar F_2$ $\bar F_8$ $S_3$
+ $\bar F_2$ $\bar F_{10}$ $S_1$ + $\bar F_2$ $H_7$ $E_5$\par
 + $s_1$ $\bar s_2$ $\chi_{14}$ + $s_2$ $\bar s_1$ $\phi _{15}$
+ $s_2$ $\bar s_2$ $\phi _{14}$ + $s_4$ $\bar s_1$ $\chi_{14}$\par
 + $s_5$ $\bar s_1$ $\phi _7$ + $s_5$ $\bar s_2$ $\phi _6$
+ $\phi _1$ $\phi _{10}$ $\phi _{15}$
+ $\phi _1$ $\phi _{11}$ $\phi _{14}$\par
 + $\phi _1$ $S_3$ $S_8$ + $\phi _2$ $\phi _7$ $\phi _{16}$
+ $\phi _2$ $\phi _9$ $\phi _{12}$
+ $\phi _3$ $\phi _6$ $\phi _{16}$\par
 + $\phi _3$ $\phi _8$ $\phi _{12}$ + $\phi _4$ $\phi _8$ $\phi _{11}$
+ $\phi _4$ $\phi _9$ $\phi _{10}$ + $\phi _5$ $S_3$ $S_5$\par
 + $\phi _6$ $\chi_{10}$ $\chi_{13}$
+ $\phi _7$ $\chi_{10}$ $\chi_{12}$
+ $\phi _8$ $\chi_7$ $\chi_{13}$ + $\phi _9$ $\chi_7$ $\chi_{12}$\par
 + $\phi _{13}$ $S_1$ $S_5$ + $\phi _{14}$ $\chi_1$ $\chi_{13}$
+ $\phi _{15}$ $\chi_1$ $\chi_{12}$ + $\phi _{16}$ $S_1$ $S_4$\par
 + $E_1$ $E_3$ $S_7$ + $E_2$ $E_3$ $S_6$ + $E_3$ $E_5$ $S_5$
+ $E_5$ $\bar E_1$ $\chi_{11}$\par
 + $S_1$ $S_2$ $\chi_{10}$ + $S_2$ $S_3$ $\chi_1$\par
}
\vskip 1in
\centerline{TABLE 4.1}
\vfil\eject

{
$W_4$ = $Q_2$ $d^c_3$ $h_3$ $\phi _{15}$
+ $Q_2$ $h_3$ $\bar t$ $\phi _{14}$
+ $Q_3$ $d^c_3$ $h_1$ $\chi_{12}$
+ $u^c_2$ $d^c_3$ $\bar t$ $\phi _4$\par
 + $d^c_1$ $d^c_2$ $\bar t$ $\bar s_2$
+ $L_3$ $e^c_3$ $h_1$ $\chi_{12}$
+ $L_3$ $\bar h_2$ $\phi _5$ $\phi _{11}$
+ $L_3$ $h_2$ $s_2$ $\phi _{12}$\par
 + $e^c_1$ $\bar s_4$ $\phi _9$ $\chi_{13}$
+ $e^c_3$ $\bar s_5$ $\phi _8$ $\chi_{12}$
+ $\bar h_2$ $h_2$ $\phi _5$ $\phi _{10}$
+ $\bar h_2$ $h_3$ $F_2$ $\bar F_1$\par
 + $\bar h_2$ $h_4$ $\phi _6$ $\phi _{11}$
+ $\bar h_2$ $h_4$ $\phi _7$ $\phi _{10}$
+ $\bar h_2$ $h_4$ $S_1$ $S_8$
+ $\bar h_4$ $h_3$ $\phi _2$ $\phi _{15}$\par
 + $\bar h_4$ $h_3$ $\phi _3$ $\phi _{14}$
+ $\bar h_4$ $h_3$ $S_3$ $S_4$
+ $h_2$ $D_2$ $\bar F_1$ $E_4$
+ $h_2$ $D_4$ $\bar F_2$ $E_4$\par
 + $h_4$ $D_1$ $\bar F_4$ $E_3$
+ $h_4$ $D_2$ $\bar F_3$ $E_3$
+ $t$ $\bar t$ $S_2$ $S_7$
+ $t$ $\bar t$ $\chi_9$ $\chi_{13}$\par
 + $D_1$ $D_3$ $T_2$ $\chi_3$
+ $D_2$ $D_4$ $S_4$ $S_8$
+ $D_3$ $D_4$ $\phi _4$ $T_3$
+ $D_3$ $D_4$ $\phi _{12}$ $T_1$\par
 + $F_1$ $\bar F_7$ $s_5$ $\chi_{11}$
+ $F_1$ $\bar F_9$ $s_2$ $\chi_{11}$
+ $\bar F_1$ $\bar F_4$ $\bar s_6$ $S_7$
+ $\bar F_2$ $\bar F_6$ $\bar s_6$ $S_7$\par
 + $H_1$ $H_7$ $S_3$ $S_4$
+ $H_2$ $H_5$ $S_3$ $S_4$
+ $s_1$ $\bar s_1$ $S_1$ $S_3$
+ $s_4$ $\bar s_4$ $\chi_6$ $\chi_{13}$\par
 + $s_4$ $\bar s_6$ $S_5$ $S_7$
+ $s_5$ $\bar s_6$ $T_1$ $T_2$
+ $\phi _2$ $\phi _{15}$ $E_3$ $\bar E_1$
+ $\phi _3$ $\phi _9$ $S_2$ $S_8$\par
 + $\phi _3$ $\phi _{14}$ $E_3$ $\bar E_1$
+ $\phi _9$ $\phi _{11}$ $S_2$ $S_4$
+ $\phi _9$ $S_2$ $S_7$ $\chi_2$
+ $\phi _9$ $S_4$ $S_8$ $\chi_{13}$\par
 + $\phi _9$ $\chi_2$ $\chi_9$ $\chi_{13}$
+ $E_2$ $E_4$ $S_7$ $\chi_{11}$
+ $E_3$ $E_4$ $S_3$ $T_1$
+ $E_3$ $\bar E_1$ $S_3$ $S_4$\par
 + $S_2$ $S_7$ $\chi_5$ $\chi_{11}$
+ $\chi_5$ $\chi_9$ $\chi_{11}$ $\chi_{13}$\par
}
\vskip 1in
\centerline{TABLE 4.2}
\vfil\eject
{\baselineskip=10.5pt\ninepoint
$W_5$ = $Q_1$ $\bar t$ $L_3$ $\phi _4$ $\varphi_4$
+ $Q_1$ $d^c_3$ $h_2$ $\phi _4$ $\varphi_4$
+ $Q_2$ $\bar t$ $L_3$ $\phi _{12}$ $\varphi_4$
+ $Q_2$ $\bar t$ $D_1$ $\bar F_9$ $E_3$\par
 + $Q_2$ $d^c_3$ $h_2$ $\phi _{12}$ $\varphi_4$
+ $Q_2$ $d^c_3$ $D_2$ $\bar F_9$ $E_3$
+ $Q_3$ $u^c_3$ $L_1$ $\bar h$ $\bar h$
+ $Q_3$ $d^c_2$ $L_1$ $\bar h$ $h_3$\par
 + $u^c_2$ $\bar t$ $\bar t$ $S_2$ $S_4$
+ $L_1$ $h$ $e^c_2$ $\bar h$ $h_3$
+ $L_1$ $h_4$ $s_3$ $\bar h$ $h_3$
+ $L_1$ $\bar h_2$ $\bar F_1$ $H_4$ $E_3$\par
 + $L_1$ $\bar h_2$ $\bar F_3$ $H_2$ $E_3$
+ $L_1$ $\bar h$ $\phi _2$ $\phi _9$ $\phi _{12}$
+ $L_1$ $\bar h$ $\phi _3$ $\phi _8$ $\phi _{12}$
+ $L_1$ $\bar h$ $E_5$ $\bar E_1$ $\varphi_1$\par
 + $L_1$ $\bar h$ $S_1$ $S_2$ $\chi_{10}$
+ $h$ $e^c_2$ $h_3$ $\phi _1$ $\phi _{16}$
+ $h$ $s_3$ $\bar h_2$ $\bar s_3$ $\chi_2$
+ $h$ $s_3$ $h_2$ $\varphi_2$ $\chi_2$\par
 + $h$ $s_3$ $L_3$ $\varphi_3$ $\chi_2$
+ $h$ $h_3$ $e^c_3$ $\phi _2$ $\phi _{14}$
+ $h$ $L_3$ $s_1$ $\phi _{10}$ $\phi _{12}$
+ $h$ $D_1$ $\bar F_5$ $E_3$ $\chi_3$\par
 + $h_4$ $\bar h_1$ $D_1$ $D_1$ $T_2$
+ $h_4$ $\bar h$ $\bar F_5$ $H_3$ $E_2$
+ $h_4$ $\bar h$ $\bar F_6$ $H_3$ $E_1$
+ $h_4$ $\bar h$ $\bar F_7$ $H_1$ $E_2$\par
 + $h_4$ $\bar h$ $\bar F_8$ $H_1$ $E_1$
+ $h_4$ $\bar h$ $\phi _8$ $E_2$ $\bar E_1$
+ $h_4$ $\bar h$ $\phi _9$ $E_1$ $\bar E_1$
+ $h_4$ $D_2$ $\bar F_3$ $\phi _9$ $E_4$\par
 + $h_4$ $D_2$ $\bar F_6$ $E_4$ $\varphi_1$
+ $e^c_1$ $\bar F_9$ $H_1$ $\phi _3$ $E_3$
+ $e^c_1$ $\bar s_3$ $\phi _7$ $S_3$ $S_5$
+ $e^c_1$ $\bar s_3$ $\phi _{15}$ $S_1$ $S_5$\par
 + $e^c_2$ $\bar h_4$ $h_3$ $\bar s_5$ $\varphi_4$
+ $e^c_2$ $D_1$ $D_4$ $\bar s_3$ $T_2$
+ $e^c_2$ $D_2$ $D_3$ $\bar s_3$ $T_2$
+ $e^c_2$ $\bar s_5$ $E_3$ $\bar E_1$ $\varphi_4$\par
 + $s_3$ $\bar h_1$ $h_2$ $\bar s_4$ $\varphi_2$
+ $s_3$ $L_2$ $h_2$ $\varphi_1$ $\varphi_2$
+ $s_3$ $\bar F_1$ $\bar F_2$ $\bar s_3$ $S_3$
+ $s_3$ $\bar F_1$ $\bar F_9$ $\bar s_3$ $S_1$\par
 + $s_3$ $\bar F_1$ $H_1$ $\bar s_3$ $E_5$
+ $s_3$ $\bar F_3$ $\bar F_4$ $\bar s_3$ $S_3$
+ $s_3$ $\bar F_3$ $\bar F_{10}$ $\bar s_3$ $S_1$
+ $s_3$ $\bar F_3$ $H_3$ $\bar s_3$ $E_5$\par
 + $s_3$ $\bar s_3$ $\phi _1$ $S_3$ $S_8$
+ $s_3$ $\bar s_3$ $\phi _5$ $S_3$ $S_5$
+ $s_3$ $\bar s_3$ $\phi _{13}$ $S_1$ $S_5$
+ $s_3$ $\bar s_3$ $\phi _{16}$ $S_1$ $S_4$\par
 + $s_3$ $\bar s_3$ $E_3$ $E_5$ $S_5$
+ $s_3$ $\bar s_5$ $\phi _8$ $\varphi_3$ $\chi_2$
+ $s_3$ $\bar s_5$ $\phi _9$ $\varphi_2$ $\chi_2$
+ $s_3$ $\bar s_5$ $\varphi_1$ $\varphi_3$ $\chi_4$\par
 + $\bar h_1$ $\bar h_2$ $\bar s_5$ $\phi _6$ $\phi _{10}$
+ $\bar h_1$ $\bar h$ $\bar F_2$ $\bar F_{10}$ $S_5$
+ $\bar h_1$ $\bar h$ $\bar F_4$ $\bar F_9$ $S_5$
+ $\bar h_1$ $L_3$ $\bar F_5$ $H_4$ $E_3$\par
 + $\bar h_1$ $L_3$ $\phi _{12}$ $S_1$ $S_6$
+ $\bar h_1$ $D_1$ $F_1$ $E_3$ $S_6$
+ $\bar h_1$ $D_1$ $\bar F_5$ $\bar s_5$ $E_3$
+ $\bar h_2$ $\bar h_4$ $\bar s_5$ $\varphi_4$ $\chi_2$\par
 + $\bar h_2$ $D_4$ $\bar F_1$ $\bar s_2$ $E_4$
+ $\bar h_2$ $D_4$ $\bar F_1$ $\bar s_4$ $E_2$
+ $\bar h$ $D_1$ $\bar F_4$ $E_2$ $\chi_{10}$
+ $\bar h$ $D_1$ $\bar F_{10}$ $E_2$ $\chi_1$\par
 + $\bar h$ $D_2$ $\bar F_4$ $E_1$ $\chi_{10}$
+ $\bar h$ $D_2$ $\bar F_{10}$ $E_1$ $\chi_1$
+ $\bar h$ $D_3$ $\bar F_2$ $E_2$ $\chi_{10}$
+ $\bar h$ $D_3$ $\bar F_9$ $E_2$ $\chi_1$\par
 + $\bar h$ $D_4$ $\bar F_2$ $E_1$ $\chi_{10}$
+ $\bar h$ $D_4$ $\bar F_9$ $E_1$ $\chi_1$
+ $\bar h_4$ $h_2$ $\phi _2$ $\phi _{12}$ $\varphi_4$
+ $\bar h_4$ $h_2$ $\phi _4$ $\phi _{10}$ $\varphi_4$\par
 + $\bar h_4$ $L_3$ $\phi _3$ $\phi _{12}$ $\varphi_4$
+ $\bar h_4$ $L_3$ $\phi _4$ $\phi _{11}$ $\varphi_4$
+ $\bar h_4$ $D_1$ $\bar F_9$ $\phi _3$ $E_3$
+ $\bar h_4$ $D_2$ $\bar F_9$ $\phi _2$ $E_3$\par
 + $h_2$ $h_2$ $s_2$ $S_2$ $S_8$
+ $h_2$ $h_3$ $\bar F_1$ $\bar F_8$ $S_2$
+ $h_2$ $h_3$ $\bar F_1$ $H_8$ $E_2$
+ $h_2$ $h_3$ $s_2$ $\phi _{14}$ $\chi_7$\par
 + $h_2$ $h_3$ $s_5$ $\phi _6$ $\chi_7$
+ $h_2$ $D_2$ $\bar F_3$ $E_3$ $\chi_8$
+ $h_2$ $D_4$ $\bar F_1$ $\bar E_1$ $S_2$
+ $h_2$ $D_4$ $H_6$ $S_2$ $S_5$\par
 + $h_3$ $L_3$ $s_2$ $\phi _{15}$ $\chi_7$
+ $h_3$ $L_3$ $s_5$ $\phi _7$ $\chi_7$
+ $h_3$ $D_3$ $H_6$ $\phi _{15}$ $T_1$
+ $h_3$ $D_4$ $\bar F_1$ $E_2$ $\chi_9$\par
 + $h_3$ $D_4$ $H_6$ $\phi _{14}$ $T_1$
+ $D_1$ $D_4$ $\phi _1$ $\phi _{13}$ $T_3$
+ $D_1$ $D_4$ $\phi _5$ $\phi _{16}$ $T_1$
+ $D_2$ $D_3$ $\phi _1$ $\phi _{13}$ $T_3$\par
 + $D_2$ $D_3$ $\phi _5$ $\phi _{16}$ $T_1$
+ $D_4$ $D_4$ $\phi _4$ $S_5$ $S_8$
+ $D_4$ $D_4$ $\phi _{12}$ $S_4$ $S_5$
+ $D_4$ $D_4$ $S_2$ $S_4$ $T_3$\par
 + $D_4$ $D_4$ $S_2$ $S_8$ $T_1$
+ $F_1$ $\bar F_1$ $\varphi_4$ $\varphi_4$ $\chi_7$
+ $F_1$ $H_1$ $s_5$ $E_3$ $S_4$
+ $F_2$ $\bar F_3$ $\varphi_4$ $\varphi_4$ $\chi_7$\par
 + $\bar F_1$ $\bar F_2$ $S_3$ $\varphi_4$ $\chi_7$
+ $\bar F_1$ $\bar F_4$ $\phi _{15}$ $S_5$ $\varphi_3$
+ $\bar F_1$ $\bar F_5$ $\bar s_3$ $\phi _7$ $S_3$
+ $\bar F_1$ $\bar F_5$ $\bar s_3$ $\phi _{15}$ $S_1$\par
 + $\bar F_1$ $\bar F_6$ $\bar s_3$ $\phi _6$ $S_3$
+ $\bar F_1$ $\bar F_6$ $\bar s_3$ $\phi _{14}$ $S_1$
+ $\bar F_1$ $\bar F_9$ $S_1$ $\varphi_4$ $\chi_7$
+ $\bar F_1$ $\bar F_{10}$ $\phi _7$ $S_5$ $\varphi_3$\par
 + $\bar F_1$ $H_1$ $\phi _3$ $\phi _{15}$ $E_4$
+ $\bar F_1$ $H_1$ $E_5$ $\varphi_4$ $\chi_7$
+ $\bar F_1$ $H_3$ $\phi _4$ $\bar E_1$ $S_3$
+ $\bar F_1$ $H_3$ $E_2$ $S_2$ $S_7$\par
 + $\bar F_1$ $H_3$ $E_2$ $\varphi_3$ $\chi_9$
+ $\bar F_3$ $\bar F_4$ $S_3$ $\varphi_4$ $\chi_7$
+ $\bar F_3$ $\bar F_7$ $\bar s_3$ $\phi _7$ $S_3$
+ $\bar F_3$ $\bar F_7$ $\bar s_3$ $\phi _{15}$ $S_1$\par
 + $\bar F_3$ $\bar F_8$ $\bar s_3$ $\phi _6$ $S_3$
+ $\bar F_3$ $\bar F_8$ $\bar s_3$ $\phi _{14}$ $S_1$
+ $\bar F_3$ $\bar F_{10}$ $S_1$ $\varphi_4$ $\chi_7$
+ $\bar F_3$ $H_3$ $\phi _3$ $\phi _{15}$ $E_4$\par
 + $\bar F_3$ $H_3$ $E_5$ $\varphi_4$ $\chi_7$
+ $\bar F_5$ $\bar F_9$ $\phi _3$ $E_3$ $E_3$
+ $\bar F_5$ $H_1$ $\phi _3$ $\phi _{13}$ $E_3$
+ $\bar F_5$ $H_4$ $\phi _8$ $E_3$ $\chi_3$\par
 + $\bar F_6$ $\bar F_9$ $\phi _2$ $E_3$ $E_3$
+ $\bar F_6$ $H_1$ $\phi _2$ $\phi _{13}$ $E_3$
+ $\bar F_9$ $H_1$ $s_4$ $\phi _2$ $E_3$
+ $\bar F_9$ $H_5$ $\phi _2$ $\phi _9$ $E_3$\par
 + $\bar F_9$ $H_5$ $\phi _3$ $\phi _8$ $E_3$
+ $H_3$ $H_6$ $\phi _4$ $S_3$ $S_5$
+ $H_3$ $H_6$ $\phi _{14}$ $\varphi_3$ $T_1$
+ $H_3$ $H_6$ $\phi _{15}$ $\varphi_2$ $T_1$\par
 + $H_3$ $H_6$ $S_2$ $S_3$ $T_1$
+ $s_2$ $s_6$ $\bar s_1$ $\bar s_5$ $\phi _6$
+ $s_2$ $\bar s_1$ $\phi _4$ $\phi _9$ $\phi _{16}$
+ $e^c_3$ $\bar s_3$ $\phi _8$ $S_2$ $S_7$\par
 + $e^c_3$ $\bar s_3$ $\phi _9$ $S_2$ $S_6$
+ $s_4$ $\bar s_3$ $\phi _6$ $S_3$ $S_5$
+ $s_4$ $\bar s_3$ $\phi _{14}$ $S_1$ $S_5$
+ $s_5$ $\bar s_1$ $\phi _1$ $\phi _9$ $\phi _{12}$\par
 + $s_6$ $\bar s_1$ $\phi _5$ $S_1$ $S_6$
+ $\phi _1$ $\phi _{16}$ $E_5$ $\bar E_1$ $\varphi_1$
+ $\phi _1$ $\phi _{16}$ $S_1$ $S_2$ $\chi_{10}$
+ $\phi _1$ $S_3$ $S_8$ $\varphi_4$ $\chi_7$\par
 + $\phi _2$ $\phi _9$ $\phi _{12}$ $\varphi_4$ $\chi_7$
+ $\phi _2$ $E_1$ $E_3$ $S_3$ $T_2$
+ $\phi _3$ $\phi _8$ $\phi _{12}$ $\varphi_4$ $\chi_7$
+ $\phi _3$ $\phi _{15}$ $E_3$ $E_4$ $S_5$\par
 + $\phi _3$ $\phi _{15}$ $E_4$ $\bar E_1$ $\varphi_1$
+ $\phi _4$ $\phi _8$ $\phi _{11}$ $\varphi_4$ $\chi_7$
+ $\phi _4$ $\phi _9$ $\phi _{10}$ $\varphi_4$ $\chi_7$
+ $\phi _4$ $S_1$ $S_8$ $\varphi_4$ $\chi_{10}$\par
 + $\phi _4$ $S_3$ $S_8$ $\varphi_4$ $\chi_1$
+ $\phi _5$ $\phi _{11}$ $S_2$ $S_7$ $\chi_6$
+ $\phi _5$ $\phi _{11}$ $\varphi_3$ $\chi_6$ $\chi_9$
+ $\phi _5$ $S_3$ $S_4$ $\varphi_4$ $T_3$\par
 + $\phi _5$ $S_3$ $S_5$ $\varphi_4$ $\chi_7$
+ $\phi _5$ $S_3$ $S_8$ $\varphi_4$ $T_1$
+ $\phi _5$ $S_7$ $S_8$ $\varphi_3$ $\chi_6$
+ $\phi _6$ $\phi _{16}$ $S_1$ $S_6$ $\chi_3$\par
 + $\phi _8$ $\phi _{12}$ $S_1$ $S_6$ $\chi_3$
+ $\phi _{12}$ $S_1$ $S_4$ $\varphi_4$ $\chi_{10}$
+ $\phi _{12}$ $S_3$ $S_4$ $\varphi_4$ $\chi_1$
+ $\phi _{13}$ $S_1$ $S_4$ $\varphi_4$ $T_3$\par
 + $\phi _{13}$ $S_1$ $S_5$ $\varphi_4$ $\chi_7$
+ $\phi _{13}$ $S_1$ $S_8$ $\varphi_4$ $T_1$
+ $\phi _{16}$ $S_1$ $S_4$ $\varphi_4$ $\chi_7$
+ $E_1$ $E_3$ $S_7$ $\varphi_4$ $\chi_7$\par
 + $E_2$ $E_3$ $S_6$ $\varphi_4$ $\chi_7$
+ $E_3$ $E_5$ $S_4$ $\varphi_4$ $T_3$
+ $E_3$ $E_5$ $S_5$ $\varphi_4$ $\chi_7$
+ $E_3$ $E_5$ $S_8$ $\varphi_4$ $T_1$\par
 + $E_3$ $\bar E_1$ $\varphi_4$ $\chi_3$ $\chi_9$
+ $E_5$ $\bar E_1$ $\varphi_1$ $\varphi_4$ $\chi_7$
+ $S_1$ $S_2$ $\varphi_4$ $\chi_7$ $\chi_{10}$
+ $S_2$ $S_3$ $\varphi_4$ $\chi_1$ $\chi_7$\par
}
%\vskip 1in
\centerline{TABLE 4.3}
\vfil\eject

{
$W_3$ = $Q_1$ $u^c_2$ $\bar h_3$ + $Q_1$ $d^c_1$ $L_2$
+ $Q_1$ $d^c_2$ $h_3$ + $Q_2$ $u^c_2$ $\bar h_2$\par
 + $Q_3$ $u^c_2$ $\bar h_1$ + $Q_3$ $d^c_3$ $h_2$
+ $Q_3$ $L_3$ $\bar t_2$ + $u^c_2$ $t_2$ $s_1$\par
 + $L_2$ $D_3$ $H_4$ + $L_2$ $D_5$ $H_3$
+ $L_3$ $D_1$ $H_{13}$ + $L_3$ $D_5$ $H_{10}$\par
 + $L_3$ $D_6$ $H_9$ + $L_3$ $D_7$ $H_7$
+ $e^c_1$ $\bar t_1$ $t_2$ + $\bar h_1$ $h_1$ $\phi _7$\par
 + $\bar h_1$ $D_5$ $\bar H_7$ + $\bar h_1$ $D_6$ $\bar H_6$
+ $\bar h_2$ $h_1$ $\phi _4$ + $\bar h_2$ $D_3$ $\bar H_7$\par
 + $\bar h_2$ $D_4$ $\bar H_6$ + $h_2$ $D_2$ $H_{13}$
+ $h_2$ $D_5$ $H_{12}$ + $h_2$ $D_6$ $H_{11}$\par
 + $h_2$ $D_8$ $H_7$ + $h_3$ $D_4$ $H_4$
+ $h_3$ $D_6$ $H_3$ + $t_1$ $\bar t_1$ $\chi_8$\par
 + $D_1$ $D_8$ $\chi_{10}$ + $D_2$ $D_7$ $\chi_{10}$
+ $D_3$ $D_6$ $\phi _6$ + $D_4$ $D_5$ $\phi _6$\par
 + $T_1$ $E_2$ $\bar F_4$ + $T_2$ $E_1$ $\bar F_4$
+ $S_1$ $\bar F_1$ $\bar F_7$
+ $S_2$ $\bar F_1$ $\bar F_6$\par
 + $S_3$ $\bar F_3$ $\bar F_5$ + $S_4$ $\bar F_2$ $\bar F_5$
+ $f_1$ $\bar F_1$ $H_{13}$ + $f_1$ $\bar F_4$ $H_7$\par
 + $f_1$ $\bar F_5$ $H_6$ + $\bar f_1$ $F_5$ $\bar H_4$
+ $\bar f_1$ $F_6$ $\bar H_3$
+ $\bar f_1$ $F_7$ $\bar H_1$\par
 + $f_2$ $\bar F_1$ $H_5$ + $f_2$ $\bar F_4$ $H_2$
+ $f_2$ $\bar F_5$ $H_1$ + $\bar f_2$ $F_5$ $\bar H_{13}$\par
 + $\bar f_2$ $F_6$ $\bar H_{12}$ + $\bar f_2$ $F_7$ $\bar H_5$
+ $f_3$ $\bar F_1$ $H_8$ + $f_3$ $\bar F_4$ $H_4$\par
 + $f_3$ $\bar F_5$ $H_3$ + $\bar f_3$ $F_5$ $\bar H_{11}$
+ $\bar f_3$ $F_6$ $\bar H_{10}$ + $\bar f_3$ $F_7$ $\bar H_2$\par
 + $F_8$ $\bar F_5$ $\bar H_9$ + $F_9$ $\bar F_5$ $\bar H_8$
+ $F_{10}$ $\bar F_1$ $\bar H_7$
+ $F_{11}$ $\bar F_1$ $\bar H_6$\par
 + $H_9$ $\bar H_{10}$ $\chi_6$ + $H_{10}$ $\bar H_{10}$ $\chi_5$
+ $H_{11}$ $\bar H_{10}$ $\chi_4$
+ $H_{12}$ $\bar H_{10}$ $\chi_3$\par
 + $s_1$ $\bar s_1$ $\chi_8$ + $T_1$ $T_2$ $\chi_{12}$
+ $T_1$ $T_3$ $\chi_9$
+ $\phi _4$ $\chi_1$ $\chi_{12}$\par
 + $\phi _7$ $\chi_2$ $\chi_9$\par
}
\vskip 1in
\centerline{TABLE 5}
\vfil\eject

\def\p{\raise2pt\hbox to5pt{\fiverm +}}
\def\m{\raise2pt\hbox to5pt{\fiverm --}}
\def\bv#1#2#3#4{V_{#1}&(#2&#3$\vert$#4)\cr}
\def\row#1#2#3#4#5#6#7#8#9{\hfill#1&\hfill#2&\hfill#3&\hfill#4
&\hfill#5&\hfill#6&\hfill#7&\hfill#8&\hfill#9\cr}

\vskip .1in
\halign to \hsbody{\kern-48truept\kern 40pt$#$:\kern .2em&#
&\kern-2.5pt$\Vert$#\cr
\bv{0}{11111111111111111111}
{1111111111111111111111111111111111111111}{1111}
\bv{1}{11100100100100100100}
{0000000000000000000000000000000000000000}{0000}
\bv{2}{00000000000000000000}
{0000000011111111000000000000000000000000}{0000}
\bv{3}{11100100010010001001}
{\p\p\p\p\p\p\p\p11110011001100000000000000000000}{0011}
\bv{4}{11001001001001100100}
{1111111100001111110000001100000000000000}{0000}
\bv{5}{11010010010010100100}
{\p\p\m\m\p\p\m\m0000\m\m\p\p00\p\p\p\p001111000000000000}{0000}
\bv{6}{11001001100100001001}
{1111000000\p\p\p\p00\p\p\p\p00000000111111110000}{0011}
\bv{7}{11010010100100001010}
{0000000000001111110000110000111100001100}{1010}
\bv{8}{00000000000000000011}
{0000000000000000000000110000110011000011}{1001}
}
\bigskip\bigskip
$$\left(\matrix{
\row{0}{0}{0}{-1/4}{0}{+1/4}{+1/4}{0}{0}
\row{0}{0}{0}{-1/2}{0}{0}{-1/2}{-1/2}{0}
\row{0}{0}{-1/2}{+1/4}{0}{0}{-1/4}{-1/2}{0}
\row{0}{0}{-1/2}{-1/4}{-1/2}{+1/4}{0}{0}{0}
\row{0}{-1/2}{-1/2}{0}{0}{-1/4}{-1/4}{0}{0}
\row{0}{-1/2}{0}{0}{0}{-1/2}{-1/2}{0}{0}
\row{0}{0}{0}{+1/4}{0}{+1/4}{0}{-1/2}{-1/2}
\row{0}{0}{0}{0}{-1/2}{-1/2}{+1/4}{0}{0}
\row{0}{0}{0}{0}{0}{0}{0}{-1/2}{-1/2}
}\right)$$

\bigskip\bigskip\bigskip\bigskip\bigskip\bigskip
\centerline{TABLE A.1}
\vfil\eject

%\bye
%\input harvmac
\def\p{\raise2pt\hbox to5pt{\fiverm +}}
\def\m{\raise2pt\hbox to5pt{\fiverm --}}
\def\bv#1#2#3#4{V_{#1}&(#2&#3$\vert$#4)\cr}
\def\row#1#2#3#4#5#6#7#8{\hfill#1&\hfill#2&\hfill#3&\hfill#4
&\hfill#5&\hfill#6&\hfill#7&\hfill#8\cr}

\vskip .1in
\halign to \hsbody{\kern-48truept\kern 40pt$#$:\kern .2em&#
&\kern-2.5pt$\Vert$#\cr
\bv{0}{11111111111111111111}
{1111111111111111111111111111111111111111}{1111}
\bv{1}{11100100100100100100}
{0000000000000000000000000000000000000000}{0000}
\bv{2}{00000000000000000000}
{0000000011111111000000000000000000000000}{0000}
\bv{3}{11100100010010001001}
{\p\p\p\p\p\p\p\p11110011001100000000000000000000}{0011}
\bv{4}{11001001001001100100}
{1111111100001111110000001100000000000000}{0000}
\bv{5}{11010010010010100100}
{\p\p\m\m\p\p\m\m0000\m\m\p\p00\p\p\p\p001111000000000000}{0000}
\bv{6}{11001001100100001001}
{1111000000\p\p\p\p00\p\p\p\p00000000111111110000}{0011}
\bv{7}{11010010100100001010}
{0000000000001111110000110000111100001100}{1010}
}
\bigskip\bigskip
$$\left(\matrix{
\row{0}{0}{0}{-1/4}{0}{+1/4}{+1/4}{0}
\row{0}{0}{0}{-1/2}{0}{0}{-1/2}{-1/2}
\row{0}{0}{-1/2}{+1/4}{0}{0}{-1/4}{-1/2}
\row{0}{0}{-1/2}{-1/4}{-1/2}{+1/4}{0}{0}
\row{0}{-1/2}{-1/2}{0}{0}{-1/4}{-1/4}{0}
\row{0}{-1/2}{0}{0}{0}{-1/2}{-1/2}{0}
\row{0}{0}{0}{+1/4}{0}{+1/4}{0}{0}
\row{0}{0}{0}{0}{-1/2}{-1/2}{-1/4}{0}
}\right)$$

\bigskip\bigskip\bigskip\bigskip\bigskip\bigskip
\centerline{TABLE A.2}
\vfil\eject

%\bye
%\input harvmac
\def\p{\raise2pt\hbox to5pt{\fiverm +}}
\def\m{\raise2pt\hbox to5pt{\fiverm --}}
\def\bv#1#2#3#4{V_{#1}&(#2&#3$\vert$#4)\cr}
\def\row#1#2#3#4#5#6#7#8{\hfill#1&\hfill#2&\hfill#3&\hfill#4
&\hfill#5&\hfill#6&\hfill#7&\hfill#8\cr}

\vskip .1in
\halign to \hsbody{\kern-48truept\kern 40pt$#$:\kern .2em&#
&\kern-2.5pt$\Vert$#\cr
\bv{0}{11111111111111111111}
{1111111111111111111111111111111111111111}{1111}
\bv{1}{11100100100100100100}
{0000000000000000000000000000000000000000}{0000}
\bv{2}{00000000000000000000}
{0000000011111111000000000000000000000000}{0000}
\bv{3}{11100100010010001001}
{\p\p\p\p\p\p\p\p11110011001100000000110011001111}{0101}
\bv{4}{11001001001001100100}
{1111111100001111110000001100000000000000}{0000}
\bv{5}{11010010010010100100}
{\p\p\m\m\p\p\m\m0000\m\m\p\p00\p\p\p\p001111000000000000}{0000}
\bv{6}{11001001100100001001}
{1111000000\p\p\p\p00\p\p\p\p00000000111111110000}{0101}
\bv{7}{11010010100100001010}
{0000000000001111110000110000111100001100}{0110}
}
\bigskip\bigskip
$$\left(\matrix{
\row{0}{0}{0}{-1/4}{0}{+1/4}{+1/4}{0}
\row{0}{0}{0}{-1/2}{0}{0}{-1/2}{-1/2}
\row{0}{0}{-1/2}{+1/4}{0}{0}{-1/4}{-1/2}
\row{0}{0}{-1/2}{+1/4}{-1/2}{+1/4}{-1/2}{-1/2}
\row{0}{-1/2}{-1/2}{0}{0}{-1/4}{-1/4}{0}
\row{0}{-1/2}{0}{0}{0}{-1/2}{-1/2}{0}
\row{0}{0}{0}{+1/4}{0}{+1/4}{0}{0}
\row{0}{0}{0}{0}{-1/2}{-1/2}{-1/4}{0}
}\right)$$

\bigskip\bigskip\bigskip\bigskip\bigskip\bigskip
\centerline{TABLE A.3}
\vfil\eject

%\bye
%\input harvmac
\def\p{\raise2pt\hbox to5pt{\fiverm +}}
\def\m{\raise2pt\hbox to5pt{\fiverm --}}
\def\bv#1#2#3#4{V_{#1}&(#2&#3$\vert$#4)\cr}
\def\row#1#2#3#4#5#6#7#8#9{\hfill#1&\hfill#2&\hfill#3&\hfill#4
&\hfill#5&\hfill#6&\hfill#7&\hfill#8&\hfill#9\cr}

\vskip .1in
\halign to \hsbody{\kern-48truept\kern 40pt$#$:\kern .2em&#
&\kern-2.5pt$\Vert$#\cr
\bv{0}{11111111111111111111}
{1111111111111111111111111111}{1111111111111111}
\bv{1}{11100100100100100100}
{0000000000000000000000000000}{0000000000000000}
\bv{2}{00000000000000000000}
{0000000011111111000000000000}{0000000000000000}
\bv{3}{11100100010010001001}
{\p\p\p\p\p\p\p\p11110011001100000011}{1111111100000000}
\bv{4}{11100100010010010010}
{1111111100111100110000001100}{0000000000000000}
\bv{5}{11010010010010100100}
{\p\p\m\m\p\p\m\m00\p\p\m\m0000\p\p\p\p111100}{0000000000000000}
\bv{6}{11001001100100001001}
{111100000000\p\p\p\p\p\p\p\p00000011}{1111000011110000}
\bv{7}{00011011011011000000}
{0000000000111100110000110000}{1100110011001100}
\bv{8}{11001001001001100100}
{1111111100111100110000110000}{1010101010101010}
}
\bigskip\bigskip
$$\left(\matrix{
\row{0}{0}{0}{-1/4}{0}{+1/4}{+1/4}{0}{0}
\row{0}{0}{0}{-1/2}{-1/2}{0}{-1/2}{0}{0}
\row{0}{0}{-1/2}{+1/4}{0}{0}{-1/4}{-1/2}{0}
\row{0}{0}{-1/2}{+1/4}{0}{+1/4}{-1/2}{-1/2}{0}
\row{0}{0}{-1/2}{0}{0}{-1/4}{0}{-1/2}{-1/2}
\row{0}{-1/2}{0}{0}{0}{-1/2}{-1/2}{-1/2}{0}
\row{0}{0}{0}{+1/4}{0}{+1/4}{0}{0}{0}
\row{0}{0}{0}{0}{0}{+1/4}{0}{-1/2}{0}
\row{0}{-1/2}{-1/2}{0}{0}{-1/4}{+1/4}{0}{-1/2}
}\right)$$

\bigskip\bigskip\bigskip\bigskip\bigskip\bigskip
\centerline{TABLE A.4}
\vfil\eject

%\bye
%\input harvmac
\def\p{\raise2pt\hbox to5pt{\fiverm +}}
\def\m{\raise2pt\hbox to5pt{\fiverm --}}
\def\bv#1#2#3#4{V_{#1}&(#2&#3$\vert$#4)\cr}
\def\row#1#2#3#4#5#6#7#8#9{\hfill#1&\hfill#2&\hfill#3&\hfill#4
&\hfill#5&\hfill#6&\hfill#7&\hfill#8&\hfill#9\cr}

\vskip .1in
\halign to \hsbody{\kern-48truept\kern 40pt$#$:\kern .2em&#
&\kern-2.5pt$\Vert$#\cr
\bv{0}{11111111111111111111}
{1111111111111111111111111111}{1111111111111111}
\bv{1}{11100100100100100100}
{0000000000000000000000000000}{0000000000000000}
\bv{2}{00000000000000000000}
{0000000011111111000000000000}{0000000000000000}
\bv{3}{11100100010010001001}
{\p\p\p\p\p\p\p\p11110011001100000011}{1111111100000000}
\bv{4}{11100100010010010010}
{1111111100111100110000001100}{0000000000000000}
\bv{5}{11001001100100001001}
{111100000000\p\p\p\p\p\p\p\p00000011}{0000000000000000}
\bv{6}{11010010010010100100}
{\p\p\m\m\p\p\m\m00\p\p\m\m0000\p\p\p\p111100}{1111000011110000}
\bv{7}{00011011011011000000}
{0000000000111100110000110000}{1100110011001100}
\bv{8}{11001001001001100100}
{1111111100111100110000110000}{1010101010101010}
}
\bigskip\bigskip
$$\left(\matrix{
\row{0}{0}{0}{-1/4}{0}{+1/4}{+1/4}{0}{0}
\row{0}{0}{0}{-1/2}{-1/2}{-1/2}{0}{0}{0}
\row{0}{0}{-1/2}{+1/4}{0}{-1/4}{0}{-1/2}{0}
\row{0}{0}{-1/2}{+1/4}{0}{0}{-1/4}{-1/2}{0}
\row{0}{0}{-1/2}{0}{0}{0}{-1/4}{-1/2}{-1/2}
\row{0}{0}{0}{+1/4}{0}{-1/2}{+1/4}{-1/2}{-1/2}
\row{0}{-1/2}{0}{0}{0}{-1/2}{0}{0}{-1/2}
\row{0}{0}{0}{0}{0}{0}{+1/4}{-1/2}{0}
\row{0}{-1/2}{-1/2}{0}{0}{+1/4}{-1/4}{0}{-1/2}
}\right)$$

\bigskip\bigskip\bigskip\bigskip\bigskip\bigskip
\centerline{TABLE A.5}
\vfil\eject

%\bye
%\input harvmac
\def\p{\raise2pt\hbox to5pt{\fiverm +}}
\def\m{\raise2pt\hbox to5pt{\fiverm --}}
\def\bv#1#2#3#4{V_{#1}&(#2&#3$\vert$#4)\cr}
\def\row#1#2#3#4#5#6#7#8#9{\hfill#1&\hfill#2&\hfill#3&\hfill#4
&\hfill#5&\hfill#6&\hfill#7&\hfill#8&\hfill#9\cr}

\vskip .1in
\halign to \hsbody{\kern-48truept\kern 40pt$#$:\kern .2em&#
&\kern-2.5pt$\Vert$#\cr
\bv{0}{11111111111111111111}
{11111111111111111111111111111111}{111111111111}
\bv{1}{11100100100100100100}
{00000000000000000000000000000000}{000000000000}
\bv{2}{00000000000000000000}
{00000000111111110000000000000000}{000000000000}
\bv{3}{11100100010010001001}
{\p\p\p\p\p\p\p\p111100110011000000001111}{111111000000}
\bv{4}{11001001001001100100}
{11111111000011111100000011000000}{000000000000}
\bv{5}{11010010010010100100}
{\p\p\m\m\p\p\m\m0000\m\m\p\p00\p\p\p\p0011110000}{000000000000}
\bv{6}{11001001100100001001}
{1111000000\p\p\p\p00\p\p\p\p000000001111}{110000111100}
\bv{7}{11010010100100001010}
{00000000000011111100001100001100}{101100110010}
\bv{8}{00000000000000000011}
{00000000000000000000001100000011}{101010101001}
}
\bigskip\bigskip
$$\left(\matrix{
\row{0}{0}{0}{-1/4}{0}{+1/4}{+1/4}{0}{0}
\row{0}{0}{0}{-1/2}{0}{0}{-1/2}{-1/2}{0}
\row{0}{0}{-1/2}{+1/4}{0}{0}{-1/4}{-1/2}{0}
\row{0}{0}{-1/2}{+1/4}{-1/2}{+1/4}{-1/2}{-1/2}{-1/2}
\row{0}{-1/2}{-1/2}{0}{0}{-1/4}{-1/4}{0}{0}
\row{0}{-1/2}{0}{0}{0}{-1/2}{-1/2}{0}{0}
\row{0}{0}{0}{+1/4}{0}{+1/4}{0}{-1/2}{-1/2}
\row{0}{0}{0}{0}{-1/2}{-1/2}{+1/4}{0}{-1/2}
\row{0}{0}{0}{0}{0}{0}{0}{0}{-1/2}
}\right)$$

\bigskip\bigskip\bigskip\bigskip\bigskip\bigskip
\centerline{TABLE A.6}
\vfil\eject

%\bye
%\input harvmac
\def\p{\raise2pt\hbox to5pt{\fiverm +}}
\def\m{\raise2pt\hbox to5pt{\fiverm --}}
\def\bv#1#2#3#4{V_{#1}&(#2&#3$\vert$#4)\cr}
\def\row#1#2#3#4#5#6#7#8#9{\hfill#1&\hfill#2&\hfill#3&\hfill#4
&\hfill#5&\hfill#6&\hfill#7&\hfill#8&\hfill#9\cr}

\vskip .1in
\halign to \hsbody{\kern-48truept\kern 40pt$#$:\kern .2em&#
&\kern-2.5pt$\Vert$#\cr
\bv{0}{11111111111111111111}
{11111111111111111111111111111111}{111111111111}
\bv{1}{11100100100100100100}
{00000000000000000000000000000000}{000000000000}
\bv{2}{00000000000000000000}
{00000000111111110000000000000000}{000000000000}
\bv{3}{11100100010010001001}
{\p\p\p\p\p\p\p\p111100110011000000001111}{111111000000}
\bv{4}{11001001001001100100}
{11111111000011111100000011000000}{000000000000}
\bv{5}{11010010010010100100}
{\p\p\m\m\p\p\m\m0000\m\m\p\p00\p\p\p\p0011110000}{000000000000}
\bv{6}{11001001100100001001}
{1111000000\p\p\p\p00\p\p\p\p000000001111}{110000111100}
\bv{7}{00000000000000000011}
{00000000000011111100000011001111}{101010101001}
\bv{8}{11010010100100001010}
{00000000000000000000001111111111}{101100110010}
}
\bigskip\bigskip
$$\left(\matrix{
\row{0}{0}{0}{-1/4}{0}{+1/4}{+1/4}{0}{0}
\row{0}{0}{0}{-1/2}{0}{0}{-1/2}{0}{-1/2}
\row{0}{0}{-1/2}{+1/4}{0}{0}{-1/4}{-1/2}{0}
\row{0}{0}{-1/2}{+1/4}{-1/2}{+1/4}{-1/2}{0}{-1/2}
\row{0}{-1/2}{-1/2}{0}{0}{-1/4}{-1/4}{-1/2}{0}
\row{0}{-1/2}{0}{0}{0}{-1/2}{-1/2}{-1/2}{-1/2}
\row{0}{0}{0}{+1/4}{0}{+1/4}{0}{-1/2}{0}
\row{0}{0}{0}{0}{-1/2}{-1/4}{0}{0}{0}
\row{0}{0}{0}{0}{0}{-1/2}{+1/4}{0}{0}
}\right)$$

\bigskip\bigskip\bigskip\bigskip\bigskip\bigskip
\centerline{TABLE A.7}
\vfil\eject

%\bye
%\input harvmac
\def\p{\raise2pt\hbox to5pt{\fiverm +}}
\def\m{\raise2pt\hbox to5pt{\fiverm --}}
\def\x{\hbox to5pt{\tenrm t}}
\def\w{\hbox to5pt{\tenrm e}}
\def\W{\hbox to5pt{\tenrm x}}
\def\bv#1#2#3#4{V_{#1}&(#2&#3$\vert$#4)\cr}
\def\row#1#2#3#4#5#6#7#8#9{\hfill#1&\hfill#2&\hfill#3&\hfill#4
&\hfill#5&\hfill#6&\hfill#7&\hfill#8&\hfill#9\cr}

\vskip .1in
\halign to \hsbody{\kern-48truept\kern 40pt$#$:\kern .2em&#
&\kern-2.5pt$\Vert$#\cr
\bv{0}{11111111111111111111}
{1111111111111111111111111111}{1111111111111111}
\bv{1}{11100100100100100100}
{0000000000000000000000000000}{0000000000000000}
\bv{2}{00000000000000000000}
{0000000011111111000000000000}{0000000000000000}
\bv{3}{11100100010010010010}
{1111111100111100110000001100}{0000000000000000}
\bv{4}{11001001100100001001}
{111100000000\p\p\p\p\p\p\p\p00000011}{0000000000000000}
\bv{5}{11010010010010100100}
{\x\x\w\w\w\w\W\W00\p\p\m\m0000\p\p\p\p\p\p1100}{1111111100000000}
\bv{6}{11001001001001100100}
{1111111100111100110000110000}{1111000011110000}
\bv{7}{00000000000000011011}
{0000000000000000000000001111}{1100110011001100}
\bv{8}{00011011011011000000}
{0000000000111100110000110000}{1010101010101010}
}
\bigskip\bigskip
$$\left(\matrix{
\row{0}{0}{0}{0}{+1/4}{-3/8}{0}{0}{0}
\row{0}{0}{0}{-1/2}{-1/2}{0}{0}{0}{0}
\row{0}{0}{-1/2}{0}{-1/4}{0}{0}{-1/2}{-1/2}
\row{0}{0}{-1/2}{0}{-1/2}{-1/2}{-1/2}{-1/2}{-1/2}
\row{0}{0}{0}{-1/2}{-1/2}{0}{-1/2}{0}{0}
\row{0}{-1/2}{0}{0}{-1/2}{-1/8}{-1/2}{-1/2}{0}
\row{0}{-1/2}{-1/2}{0}{+1/4}{+1/8}{-1/2}{-1/2}{0}
\row{0}{0}{-1/2}{-1/2}{0}{+1/4}{0}{-1/2}{-1/2}
\row{0}{0}{0}{0}{-1/2}{-1/8}{0}{0}{-1/2}
}\right)$$

\bigskip\bigskip\bigskip\bigskip\bigskip\bigskip
\centerline{TABLE A.8}

\bye